\documentclass[longauth]{aa} 

\usepackage{graphicx}
\usepackage{txfonts}
\usepackage{amssymb}
\usepackage{amsmath}
\usepackage{graphicx}
\usepackage[colorlinks=true, allcolors=blue]{hyperref}
\usepackage{booktabs}
\usepackage{amsfonts}     
\usepackage[T1]{fontenc} 
\usepackage[utf8]{inputenc} 
\usepackage{multicol}
\usepackage{xcolor}
\usepackage{soul}
\usepackage{rotating}
\usepackage{comment}
\usepackage{multirow}
\usepackage{longtable}
\usepackage{lscape}
\usepackage{color}
\usepackage{url}

\usepackage{orcidlink}
\usepackage{academicons}

\linenumbers

\begin{document}

\title{Testing particle acceleration in blazar jets with continuous high-cadence optical polarization observations}


\author{
Ioannis Liodakis, \inst{\ref{AstroCrete},\ref{NASA_Alabama}} \orcid{0000-0001-9200-4006} 
Sebastian Kiehlmann, \inst{\ref{AstroCrete},\ref{AstroCrete2}} \orcid{0000-0001-6314-9177} 
Alan P. Marscher, \inst{\ref{BostonUni}} \orcid{0000-0001-7396-3332} 
Haocheng Zhang, \inst{\ref{Maryland},\ref{NASA_goddard}} \orcid{0000-0001-9826-1759}
Dmitry Blinov, \inst{\ref{AstroCrete},\ref{AstroCrete2}} 
Svetlana G. Jorstad, \inst{\ref{BostonUni},\ref{StPetersburg}} \orcid{0000-0001-6158-1708} 
Iv\'{a}n Agudo, \inst{\ref{InstAstro_Granada}} \orcid{0000-0002-3777-6182} 
Erika Ben\'itez, \inst{\ref{OAN-SPM}}  \orcid{000-0003-1018-2613} 
Andrei Berdyugin, \inst{\ref{UTU}} 
Giacomo Bonnoli, \inst{\ref{INAF_Merate},\ref{InstAstro_Granada}} \orcid{0000-0003-2464-9077} 
Carolina Casadio, \inst{\ref{AstroCrete}} 
Chien-Ting Chen, \inst{\ref{NASA_Alabama}}
Wen-Ping Chen, \inst{\ref{Lulin}}
Steven R. Ehlert, \inst{\ref{NASA_Alabama}} \orcid{0000-0003-4420-2838} 
Juan Escudero, \inst{\ref{InstAstro_Granada}} 
Tatiana S. Grishina, \inst{\ref{StPetersburg}} \orcid{0000-0002-3953-6676} 
David Hiriart, \inst{\ref{OAN-SPM}} \orcid{000-0002-4711-7658}
Angela Hsu,\inst{\ref{Lulin}}
Ryo Imazawa, \inst{\ref{Jap_UniHiroshima}} 
Helen E. Jermak, \inst{\ref{LJMU}}\orcid{0000-0002-1197-8501} 
Jincen Jose, \inst{\ref{ARIES},\ref{ARIES2}} 
Philip Kaaret, \inst{\ref{NASA_Alabama}} \orcid{0000-0002-3638-0637} 
Evgenia N. Kopatskaya, \inst{\ref{StPetersburg}} \orcid{0000-0001-9518-337X} 
Bhavana Lalchand,\inst{\ref{Lulin}}
Elena G. Larionova, \inst{\ref{StPetersburg}} \orcid{0000-0002-2471-6500} 
Elina Lindfors, \inst{\ref{UTU},\ref{FINCA}} 
Jos\'{e} M. L\'{o}pez,\inst{\ref{OAN-SPM2}} 
Callum McCall, \inst{\ref{LJMU}}\orcid{0000-0002-3375-3397} 
Daria A. Morozova, \inst{\ref{StPetersburg}} \orcid{0000-0002-9407-7804} 
Efthymios Palaiologou, \inst{\ref{AstroCrete}} 
Shivangi Pandey, \inst{\ref{ARIES},\ref{ARIES3}} 
Juri Poutanen, \inst{\ref{UTU}} \orcid{0000-0002-0983-0049} 
Suvendu Rakshit, \inst{\ref{ARIES}} 
Pablo Reig, \inst{\ref{AstroCrete}} 
Mahito Sasada, \inst{\ref{Jap_TokyoInstTech}} \orcid{0000-0001-5946-9960} 
Sergey S. Savchenko, \inst{\ref{StPetersburg},\ref{Pulkovo}} \orcid{0000-0003-4147-3851} 
Elena Shablovinskaya, \inst{\ref{SAO},\ref{Chile}} 
Sharma Neha, \inst{\ref{ARIES}} 
Manisha Shrestha, \inst{\ref{Steward}} \orcid{0000-0002-4022-1874}
Iain A. Steele, \inst{\ref{LJMU}} 
Ivan S. Troitskiy, \inst{\ref{StPetersburg}} \orcid{0000-0002-4218-0148} 
Yulia V. Troitskaya, \inst{\ref{StPetersburg}} \orcid{0000-0002-9907-9876} 
Makoto Uemura, \inst{\ref{Jap_UniHiroshima},\ref{Jap_ASCHiroshima},\ref{Jap_CoreUHiroshima}} 
Andrey A. Vasilyev, \inst{\ref{StPetersburg}} \orcid{0000-0002-8293-0214} 
Zachary Weaver, \inst{\ref{BostonUni}}\orcid{0000-0001-6314-0690} 
Klaas Wiersema, \inst{\ref{Herfordshire},\ref{Leicester}} \orcid{0000-0002-9133-7957} 
Martin C. Weisskopf \inst{\ref{NASA_Alabama}} \orcid{0000-0002-5270-4240} 
}
\institute{
Institute of Astrophysics, Foundation for Research and Technology-Hellas, GR-70013 Heraklion, Greece \label{AstroCrete} 
\and
NASA Marshall Space Flight Center, Huntsville, AL 35812, USA \label{NASA_Alabama}
\and
Department of Physics, University of Crete, GR-70013 Heraklion, Greece \label{AstroCrete2} 
\and
Institute for Astrophysical Research, Boston University, 725 Commonwealth Avenue, Boston, MA 02215, USA \label{BostonUni}
\and
University of Maryland Baltimore County Baltimore, MD 21250, USA\label{Maryland}
\and
NASA Goddard Space Flight Center Greenbelt, MD 20771, USA \label{NASA_goddard}
\and
Saint Petersburg State University, 7/9 Universitetskaya nab., St. Petersburg, 199034 Russia \label{StPetersburg}
\and
Instituto de Astrof\'{i}sica de Andaluc\'{i}a, IAA-CSIC, Glorieta de la Astronom\'{i}a s/n, 18008 Granada, Spain \label{InstAstro_Granada}
\and
Instituto de Astronom\'{i}a, Universidad Nacional Aut\'{o}noma de M\'{e}xico, A.P. 70-264, CDMX 04510, Mexico, Ensenada 22800, Baja California, M\'{e}xico\label{OAN-SPM}
\and
Department of Physics and Astronomy, University of Turku, FI-20014, Finland \label{UTU}
\and
INAF Osservatorio Astronomico di Brera, Via E. Bianchi 46, 23807 Merate (LC), Italy \label{INAF_Merate}
\and
Institute of Astronomy, National Central University, Taoyuan 32001, Taiwan \label{Lulin}
\and
Department of Physics, Graduate School of Advanced Science and Engineering, Hiroshima University Kagamiyama, 1-3-1 Higashi-Hiroshima, Hiroshima 739-8526, Japan \label{Jap_UniHiroshima}
\and
Astrophysics Research Institute, Liverpool John Moores University, Liverpool Science Park IC2, 146 Brownlow Hill, Liverpool, UK, L3 5RF \label{LJMU}
\and
Finnish Centre for Astronomy with ESO (FINCA), FI-20014 University of Turku, Finland \label{FINCA}
\and
Facultad de Ciencias, Universidad Autónoma de Baja California, Campus El Sauzal, 22800 Ensenada, Baja California, Mexico\label{OAN-SPM2}
\and
Aryabhatta  Research Institute of Observational Sciences (ARIES), Manora Peak, Nainital, 263002 India \label{ARIES}
\and
Center for Basic Sciences, Pt. Ravishankar Shukla University, Raipur, Chhattisgarh, 492010, India \label{ARIES2}
\and
Department of Applied Physics/Physics, M.J.P. Rohilkhand University, Bareilly, Uttar Pradesh - 24300 \label{ARIES3}
\and
Institute of Innovative Research (IIR), Tokyo Institute of Technology, 2-12-1 Ookayama, Meguro-ku, Tokyo 152-8551, Japan \label{Jap_TokyoInstTech}
\and
Pulkovo Observatory, St.Petersburg, 196140, Russia \label{Pulkovo}
\and
Special astrophysical observatory of Russian Academy of Sciences, Nizhnĳ Arkhyz, Karachai-Cherkessian Republic, 369167, Russia\label{SAO}
\and
Instituto de Estudios Astrof\'isicos, Facultad de Ingenier\'ia y Ciencias, Universidad Diego Portales, Santiago, Regi\'on Metropolitana, 8370191 Chile\label{Chile}
\and
Steward Observatory, University of Arizona, 933 North Cherry Avenue, Tucson, AZ 85721-0065, USA \label{Steward}
\and
Hiroshima Astrophysical Science Center, Hiroshima University 1-3-1 Kagamiyama, Higashi-Hiroshima, Hiroshima 739-8526, Japan \label{Jap_ASCHiroshima}
\and
Core Research for Energetic Universe (Core-U), Hiroshima University, 1-3-1 Kagamiyama, Higashi-Hiroshima, Hiroshima 739-8526, Japan \label{Jap_CoreUHiroshima}
\and
Centre for Astrophysics Research, University of Hertfordshire, Hatfield, AL10 9AB, UK \label{Herfordshire}
\and
School of Physics and Astronomy, University of Leicester, University Road, Leicester, LE1 7RH, UK\label{Leicester}
}



\abstract{Variability can be the pathway to understanding the physical processes in astrophysical jets, however, the high-cadence observations required to test particle acceleration models are still missing. Here we report on the first attempt to produce continuous, $>24$ hour polarization light curves of blazars using telescopes distributed across the globe and the rotation of the Earth to avoid the rising Sun. Our campaign involved 16 telescopes in Asia, Europe, and North America. We observed BL Lacertae and CGRaBS~J0211+1051 for a combined 685 telescope hours. We find large variations in the polarization degree and angle for both sources in sub-hour timescales as well as a $\sim180\degr$ rotation of the polarization angle in CGRaBS~J0211+1051 in less than two days. We compared our high-cadence observations to Particle-In-Cell magnetic reconnection and turbulent plasma simulations.  We find that although the state of the art simulation frameworks can produce a large fraction of the polarization properties, they do not account for the entirety of the observed polarization behavior in blazar jets.}


\titlerunning{Testing particle acceleration in blazar jets}
\authorrunning{Liodakis et al.}

\maketitle
%
\section{Introduction} \label{sec:intro}

Blazars are active galactic nuclei with powerful jets oriented within a few degrees from the line of sight of an observer on Earth \citep{Blandford2019,Hovatta2019}. They show a plethora of exciting behavior, but are most notably known for their broadband emission from radio to $\gamma$-rays \cite[e.g.,][]{Ajello2020}, extreme variability down to minute timescales \cite[e.g.,][]{Ackermann2016} and their highly polarized emission, with the polarization degree that can exceed 45\% \cite[e.g.,][]{Shao2019} and highly relativistic jets \citep{Lister2021,Weaver2022} . The origin of the extreme, diverse variability in different wavelengths is not fully understood, but holds the keys to probing physical processes in astrophysical jets. A few models have now been proposed to explain the observed variability patterns. Those include shock-in-jets \citep{Marscher1985}, jet-in-jet \citep{Giannios2009}, turbulence \citep{Marscher2014,Webb2023}, magnetic reconnection \citep{Hosking2020,Zhang2020}, shock-shock collisions \citep{Liodakis2020}, kink instabilities \citep{Zhang2017}, Doppler factor variations \citep{Raiteri2017,Raiteri2017-II}, and others. A lot of effort has been placed in characterizing the jet's variability on diverse timescales. The recent introduction of datasets from exoplanet satellites like Kepler and TESS \citep{Sasada2017,Weaver2020,Raiteri2021} has given an additional unique view of blazar variability. However, despite all these efforts we still lack a consensus of the mechanisms that drive variability in blazar jets. 

Polarization can be an additional probe into the physical processes that govern the jets. This is because different models often predict different polarization properties \cite[e.g.,][]{Marscher2014,Zhang2014,Peirson2018,Peirson2019,Tavecchio2021}. Moreover, blazars show peculiar polarization behavior that often takes place in the form of rotations of the polarization angle \cite[e.g.,][]{Marscher2008,Marscher2010,Blinov2015,Blinov2016-II}.  Polarization studies of blazars have been limited by the cadence of the observations and the lack of dedicated experiments (see \citealp{Kiehlmann2021} for discussion) and until recently wavelength coverage, as most studies are in the optical. The recent launch of the Imaging X-ray Polarimetry Explorer (IXPE, \citealp{Weisskopf2022}) has opened new avenues to studying the high-energy Universe. The first observations of blazars suggest that particle acceleration in jets takes place in shocks with the emission becoming energy-stratified owing to particle cooling \citep{Liodakis2022,DiGesu2022,Kouch2024}. More interestingly, IXPE already managed to detect the first X-ray polarization angle rotation in Mrk~421 \citep{DiGesu2023,Kim2024} in a blind survey. Mrk~421 showed a $>$360$^\circ$ rotation with a rate of 80--90\degr\,d$^{-1}$. This would suggest that large variations of the polarization angle can take place within a single observing night (see also \citealp{MAGICCollaboration2018}) and are missed in the optical  because of the 180$^\circ$ ambiguity of the polarization angle and our inability to monitor a blazar from one location continuously. Uninterrupted observations have only been achieved in a limited number of campaigns \cite[e.g.,][]{Bhatta2016,Weaver2020}.

Here we aim to combine the power of variability and polarization by producing continuous light curves from the ground. This is achieved by combining multiple telescopes across the world using the rotation of the Earth. Different telescopes across the world have often been combined for blazar studies \cite[e.g.,][]{Raiteri2021-II} and other purposes \cite[e.g.,][]{LasCumbresObservatory2013}.Here we focus on polarimetry and probing shorter variability timescales by producing continuous, uninterrupted, longer than 24-hour time-series that we can use to test particle acceleration models. In Sect. \ref{sec:telescopes} we describe the telescopes used in this work and our observing strategy, while in Sect. \ref{sec:data} we discuss our analysis procedures. In Sect. \ref{sec:var_prop} we present the final time-series and examine their variability properties. In Sect. \ref{sec:test_models} we test different models of particle acceleration, and in Sect. \ref{sec:discussion} we discuss our findings. We focus on linear polarization, which we refer to as ``polarization'' throughout the paper for simplicity.

\section{Telescopes and observing strategy} \label{sec:telescopes}

Our campaign, which we dubbed NOPE - the NOn-stop Polarization Experiment, consisted of 16 telescopes across the world with a combined 685 telescope hours over seven nights.  Those telescopes/observatories are, the Aryabhatta Research Institute of Observational Sciences (ARIES), the Calar Alto  Observatory, the Crimean observatory, T60 at the Haleakala observatory  \citep{Piirola2014,Piirola2020}, the Kanata telescope \citep{Uemura2017}, the Liverpool Telescope \citep{Steele2004}, LX-200 \citep{Larionov2008},  Lulin observatory, the Nordic Optical Telescope \citep{Hovatta2016,Nilsson2018}, the Perkins Telescope Observatory (PTO, \citealp{Jorstad2010}), the Observatorio Astron\'{o}mico Nacional at San Pedro M\'{a}rtir (OAN-SPM), the Zeiss-1000/MAGIC at the Special Astrophysical Observatory of RAS \citep{2023RASTI...2..657A,2020AstBu..75..486K}, Sierra Nevada Observatory, RoboPol at the Skinakas observatory \citep{Ramaprakash2019}, and the University of Leicester Observatory \citep{Wiersema2023}. NOPE was scheduled for the dark time of 2--8 November 2021. The volcano eruption in La Palma prevented us from acquiring data from the Liverpool Telescope and the Nordic Optical Telescope. Weather related reasons prevented the use of the telescope at the University of Leicester Observatory as well. Additional telescope hours were lost due to weather in different locations, however, all remaining telescopes provided a reasonable amount of data.

To probe fast variability timescales we selected the most variable sources found from the four years of monitoring in the RoboPol sample  \citep{Blinov2021} that were visible for at least half the night from most locations. Since rotations of the polarization has been shown to be connected to $\gamma$-ray activity \citep{Blinov2018}, we monitored the $\gamma$-ray light curves of all the objects in the sample \citep{Baldini2021,repository2023} as well as for other flaring sources using alert brokers (e.g., Astronomer's telegram\footnote{\url{https://www.astronomerstelegram.org/}}).

\section{Optical polarization observations} \label{sec:data}

\subsection{Observations and data reduction} \label{sec:data-obs}

At the time of the scheduled observations there were no outbursts or elevated activity reported, hence we opted to observe BL Lacertae (BL Lac) which was in a prolonged outburst that had lasted a few months \citep{Raiteri2023} and CGRaBS J0211+1051 (hereafter J0211) that was in a historically low brightness and low polarization degree period \citep{Blinov2021}. BL Lac is typically a low synchrotron peaked blazar (i.e., synchrotron peak frequency $<10^{14}$ Hz, \citealp{Ajello2020,Middei2023}), but turns into an intermediate peaked source during flares (i.e., synchrotron peak frequency $\rm 10^{14}<\nu_{syn}<10^{15}$ Hz, e.g., \citealp{Peirson2023}). J0211 is an intermediate peaked blazar \citep{Peirson2022}. BL Lac was observed for the first half of each night while J0211 for the second half. The observations were performed in the R-band. The data were analyzed using either standard analysis procedures or existing pipelines at individual observatories \cite[e.g.,][]{King2014,Panopoulou2015,Nilsson2018}. All the polarimetric measurements were performed with a 5$^{\prime\prime}$ aperture radius. We pre-selected several polarized (HD 204827 \& BD +59.389) and unpolarized (BD+32.3739, BD +28.4211, HD 212311, HD 14069, G191B2B) standard stars, commonly used by blazar monitoring program like the Steward observatory\footnote{\url{http://james.as.arizona.edu/~psmith/SPOL/polstds.html}} and RoboPol \citep{Blinov2023}, to be used by all the observatories. Since we are interested in the polarization variability, we did not apply any correction to the polarization degree from the dilution of the host-galaxy which, at the selected aperture, should be negligible \cite[e.g.,][]{Meisner2010}.

For both sources we are able to achieve a median cadence of five minutes, with the shortest interval between observations to be zero, and the longest, that occurred during the end of our campaign, to be roughly nine hours for BL Lac and twelve hours for J0211.

\subsection{Data post-processing} \label{sec:data-post}

\paragraph{Systematic shifts:}
Once all the data were collected, small systematic shifts of the order of $\rm<0.1^{mag}$ and $<$0.5\% were applied to the brightness and polarization degree respectively to align overlapping observations. Those shifts originate from the different efficiency in the instruments, instrumental setup, seeing, and minor differences in the analysis pipelines. 
\paragraph{Data binning:}
We average data points in bins of 30\,min duration. These bins are interactively identified\footnote{We use the \texttt{smart\_binning} function from \url{https://github.com/skiehl/timeseriestools}.}, using the method described in \citet[][Sect.~2.1.2]{2015PhDT.......630K}. In the selected bins we calculate the mean value of $q$ and $u$. Even though the variability in the bins is below the noise level, we expect some intrinsic variability. Therefore, we do not use the uncertainty-weighted mean, which is based on the assumption that all measurements are estimates of the same intrinsic value. To derive realistic uncertainties for the average $q$ and $u$ we calculate (i) the mean of the corresponding uncertainties and (ii) the standard deviation of $q$ and $u$ in the bin and we use the larger value of the two. We then calculate the corresponding $p$ and $\chi$ using Eq.~\ref{eq:pol}, 
\begin{equation}
   p = \sqrt{q^2 + u^2},
    \chi = \frac{1}{2} \arctan \left( \frac{u}{q} \right),
    \label{eq:pol}
\end{equation}
and we use the procedure described in \citet{Blinov2021} to estimate the corresponding uncertainties. We correct for the $\pm180^\circ$ ambiguity of the polarization angle following \cite{Kiehlmann2016}. The brightness and polarization light curves for both sources used in the paper are available at the Harvard Dataverse \citep{DVN/IETSXS_2024}\footnote{\url{https://doi.org/10.7910/DVN/IETSXS}}.

\section{Variability properties}\label{sec:var_prop}

\begin{figure*}
    \centering
     \includegraphics[width=\textwidth]{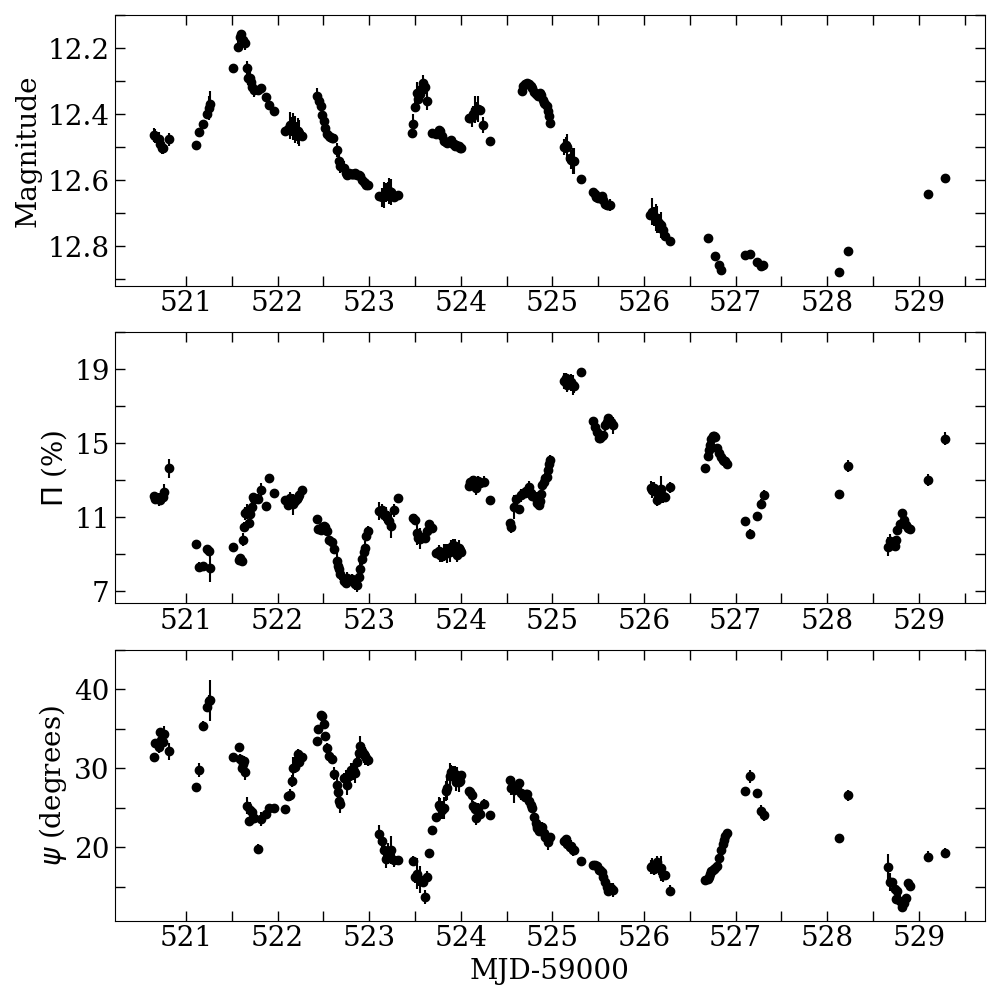}
    \caption{NOPE observations of BL Lac. The top panel shows the R-band magnitude, the middle panel the polarization degree, and the bottom panel the polarization angle. The observations have been grouped in 30min bins (see text).}
    \label{plt:BLLac}
\end{figure*}

\begin{figure*}
    \centering
     \includegraphics[width=\textwidth]{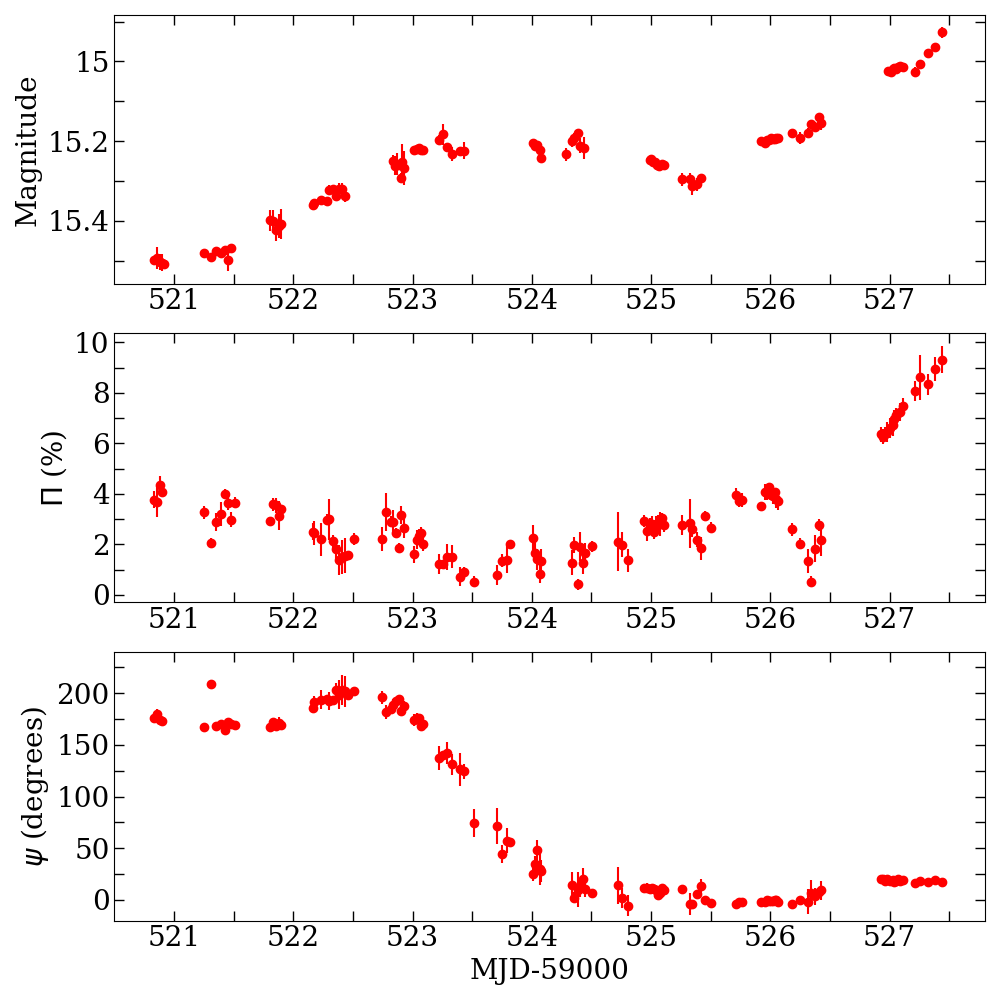}
    \caption{NOPE observations of CGRaBS~J0211+1051. The top panel shows the R-band magnitude, the middle panel the polarization degree, and the bottom panel the polarization angle. The observations have been grouped in 30min bins (see text).}
    \label{plt:J0211}
\end{figure*}

\subsection{Light curves} \label{sec:data-lcs}

\begin{figure}
    \centering
    \resizebox{\hsize}{!}{   \includegraphics[width=\textwidth]{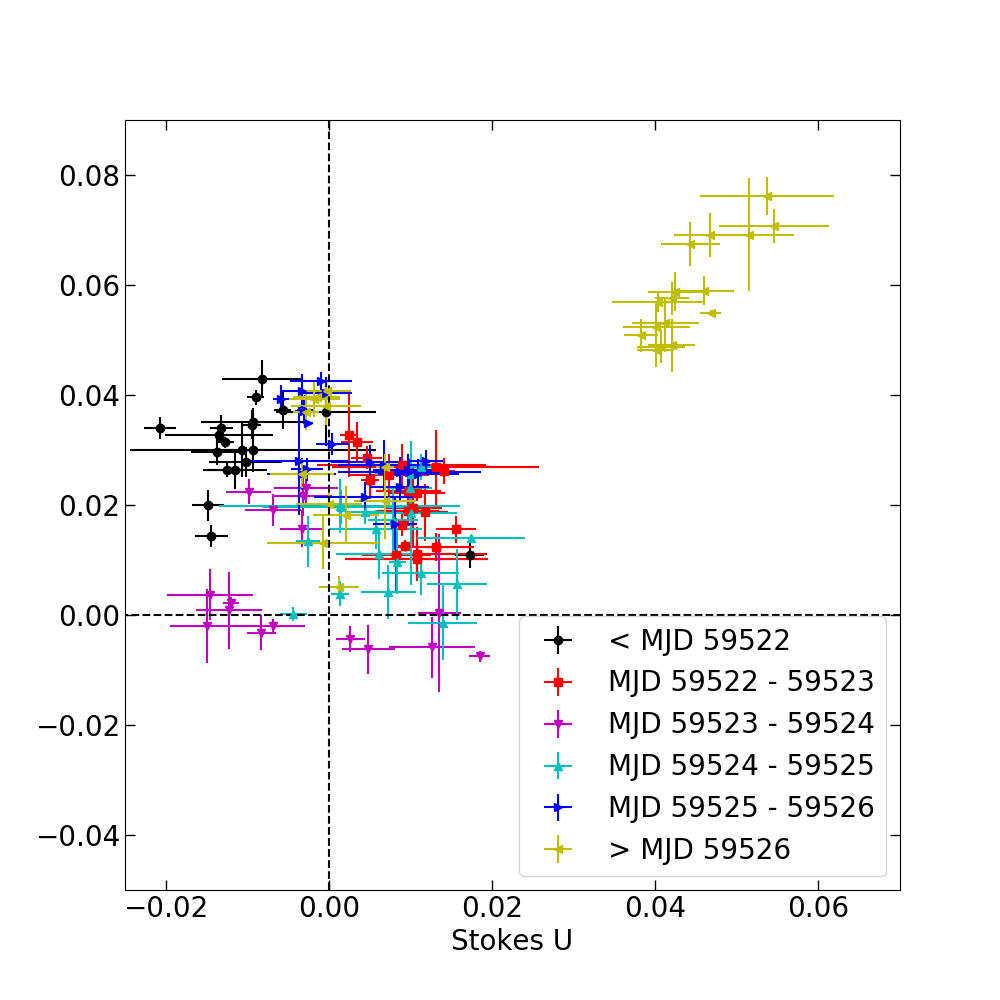}}
    \caption{Stokes Q versus Stokes U for CGRaBS~J0211+1051. The data have been split by MJD to show the progression of the rotation in the Q-U plane. The black dashed lines mark 0-0.}
    \label{plt:J0211_rotation}
\end{figure}

Figures \ref{plt:BLLac} and \ref{plt:J0211} show the observations for BL Lac and J0211 respectively. BL Lac  shows rapid intranight flaring that would be impossible to fully characterize without longer than twelve hour observations. The R-band magnitude of the source varies from a minimum of 12.87$\rm^{mag}$ to a maximum of 12.15$\rm^{mag}$, with a median of 12.48$\rm^{mag}$ and  $\rm\sim$0.5$^{\rm mag}$ flares. The polarization degree ($\Pi$) has a min, max of 7.3\% and 18.9\%, respectively, with a median of 11.7\%. The polarization angle ($\psi$) remains near the jet axis on the sky (10$^\circ\pm2^\circ$, \citealp{Weaver2022}), fluctuating from 12$^\circ$ to 39$^\circ$ with a median of 25$^\circ$. There is no correlation between brightness and polarization as is typical of blazars \citep{Ikejiri2011,Blinov2016,Jermak2016}. 

J0211 exhibits a more complex behavior with smoother variations than BL Lac. The R-band magnitude ranges from 15.5$\rm^{mag}$ to 14.9$\rm^{mag}$ with a median of 15.2$\rm^{mag}$. $\Pi$ ranges from 0.43\% to 9.3\% with a median of 2.7\%. We observe both a correlation and anti-correlation between brightness and the polarization degree. At the start of our campaign, J0211 shows a smooth increase in brightness. At the same time $\Pi$ decreases from about 5\% to the lowest observed value of 0.43\%, which occurs at peak brightness. Then the polarization degree starts to recover. During the increase in brightness and drop of $\Pi$ we observe a smooth monotonic 185$^\circ$ rotation of the $\psi$ from $\sim$200$^\circ$ to $\sim$15$^\circ$. Figure \ref{plt:J0211_rotation} shows the progression of the rotation in the Q-U plane. There is a clear circular motion of the Stokes vectors indicating the presence of a rotation slightly offset from 0-0. The jet direction has been determined at 15~GHz to be $88^\circ\pm9^\circ$ \citep{Hodge2018}, which suggests that the rotation started and ended roughly perpendicular to the jet. The drop of $\Pi$ during $\psi$ rotations in the optical is a common feature in the blazar population \citep{Blinov2016}. After the rotation, $\psi$ remains constant while the brightness and $\Pi$ show a clear correlation.

\subsection{Bayesian Block analysis}

\begin{table*}
\setlength{\tabcolsep}{11pt}
\centering
  \caption{Summary of Bayesian Block analysis. The columns list the source and corresponding average, minimum, and maximum values. The rows are for the variability time scale (h), flare amplitude (mJy for brigthness and \% for $\Pi$), and flare duration (h).}
  \label{tab:BB}
\begin{tabular}{@{}lcccc@{}}
\hline
 \hline
 & Source  & Average & minimum  & maximum \\
  \hline
Variability timescale (brightness, hours) & BL Lac & 4.14 & 0.87 & 23.61  \\
Variability timescale ($\Pi$, hours) & BL Lac & 4.78 & 0.88 & 28.61 \\
Flare amplitude (brightness, mJy) & BL Lac & 14.11 & 13.23 & 19.39\\
Flare amplitude ($\Pi$, \%) & BL Lac & 5.19 & 3.35 & 10.65\\
Flare duration (brightness, hours) & BL Lac & 25.5 & 16.66 & 77.08 \\
Flare duration ($\Pi$, hours) & BL Lac & 24.07 & 18.5 & 37.25 \\
\hline
Variability timescale (brightness, hours)  & J0211 & 11.77 & 3.53 &  23.80 \\
Variability timescale ($\Pi$, hours) & J0211 & 13.30 & 3.02 & 34.22 \\
Flare amplitude (brightness, mJy) & J0211 & 0.57 & -- &  -- \\
Flare amplitude ($\Pi$, \%) & J0211 & 2.05 & 1.37 & 2.72 \\
Flare duration (brightness, hours) & J0211 & 75.66 & -- & -- \\
Flare duration ($\Pi$, hours) & J0211 & 43.65  & 31.6 & 55.7 \\
\hline
\end{tabular}
\end{table*}

\begin{figure}
\centering
 \resizebox{\hsize}{!}{  \includegraphics[width=\textwidth]{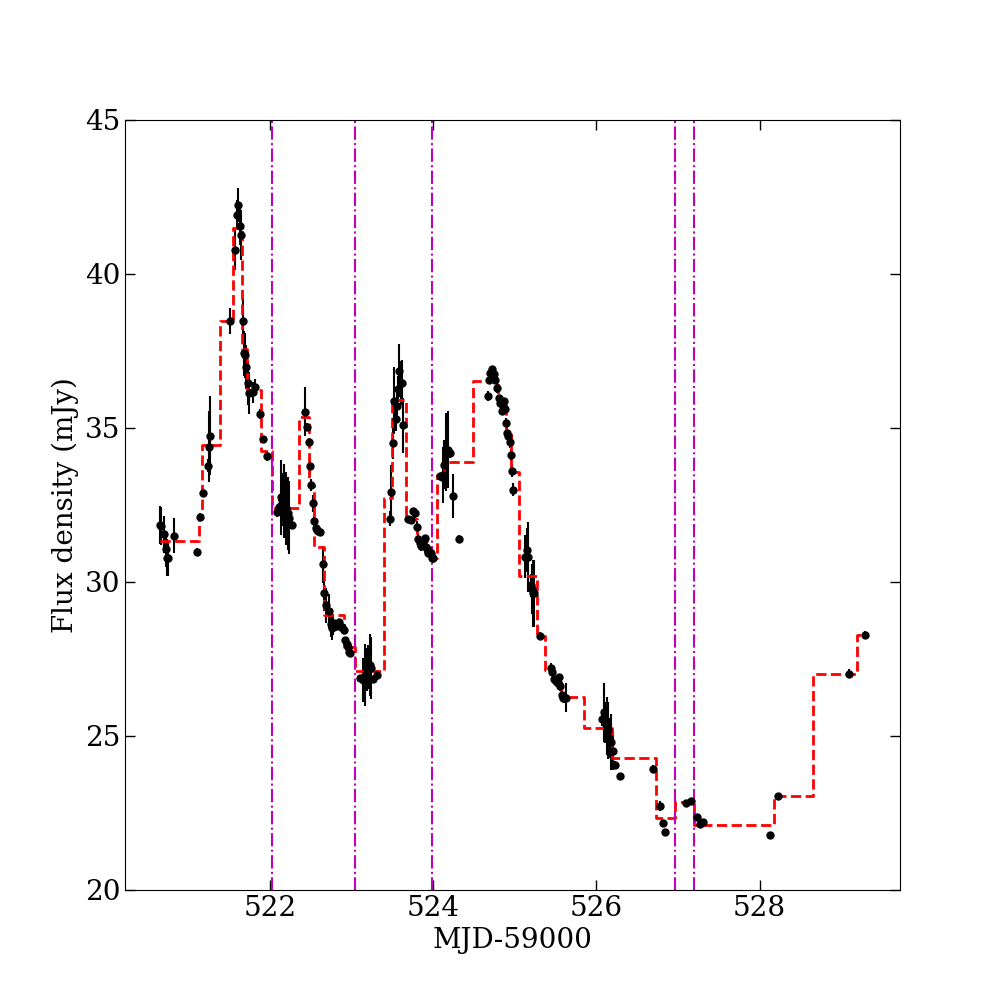}}
\resizebox{\hsize}{!}{  \includegraphics[width=\textwidth]{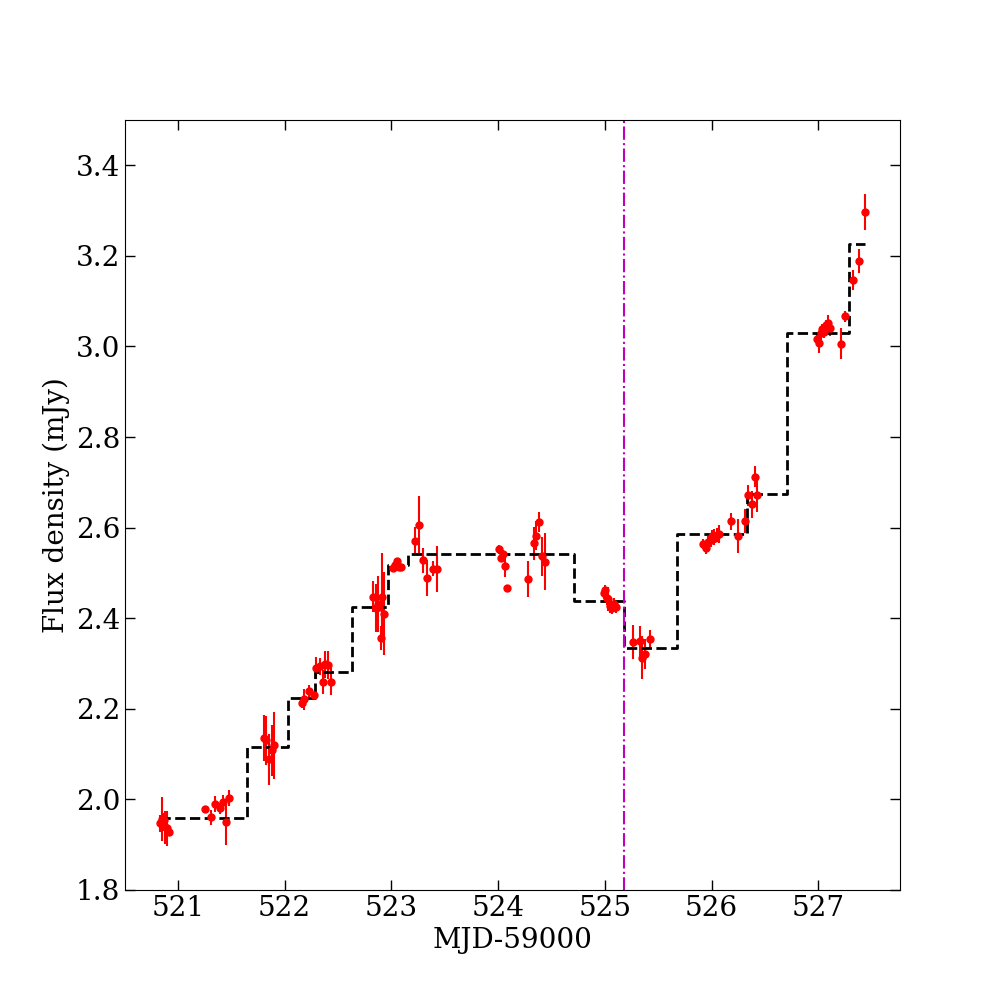}}
 \caption{Bayesian Block representation (dashed line) for the flux density variations of the BL Lac (top panel) and CGRaBS~J0211+1051 (bottom panel). The vertical magenta lines show the edges of the time-intervals identified by the HOP module.}
    \label{plt:BB_mag}
\end{figure}

\begin{figure}
\centering
 \resizebox{\hsize}{!}{  \includegraphics[width=\textwidth]{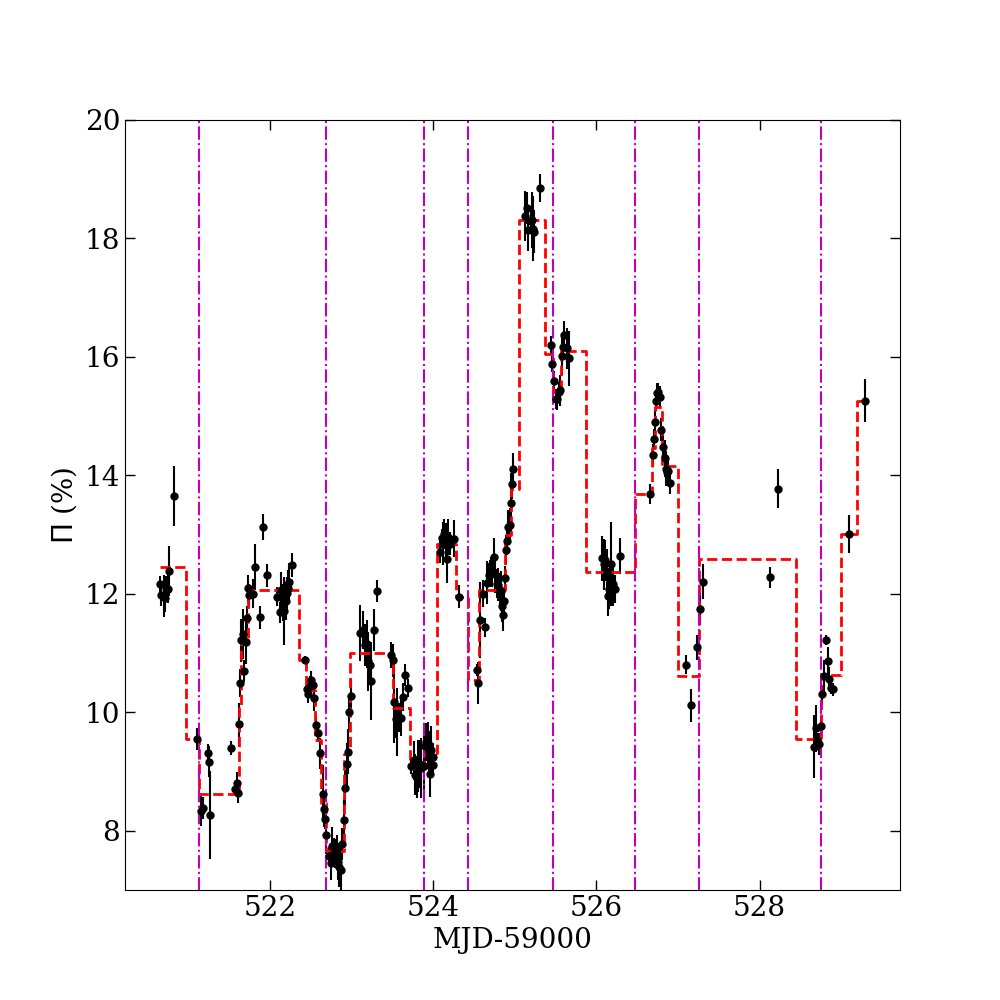}}
\resizebox{\hsize}{!}{  \includegraphics[width=\textwidth]{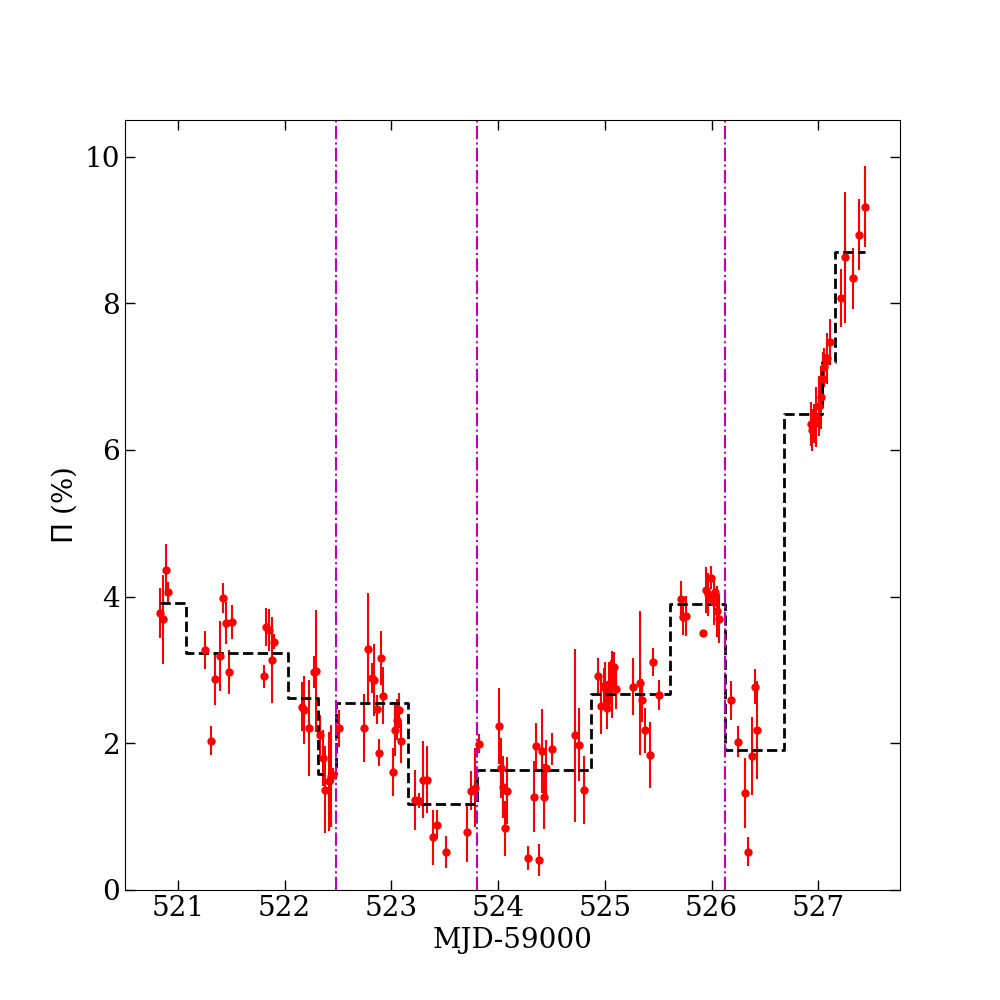}}
 \caption{Bayesian Block representation (dashed line) for the polarization degree variations of the BL Lac (top panel) and CGRaBS~J0211+1051 (bottom panel). The vertical magenta lines show the edges of the time-intervals identified by the HOP module.}
    \label{plt:BB_pol}
\end{figure}

We use Bayesian Blocks \cite[BB,][]{Scargle2013} to model the brightness and polarization degree light curves \cite[e.g.,][]{Liodakis2018,Liodakis2019}. BB has only one parameter {\it ncrprior}, related to the prior for the number of bins. Following \cite{deJaeger2023}, we use a value of the {\it ncrprior} that yields a false-positive rate ($p_0$) of $p_0=0.01$. We also implement the HOP module \citep{Eisenstein1998,Meyer2019} that allows us to trace changes in the block flux derivative. That allows us to separate extended time-intervals that could be considered as flares or parts of a flare. For the brightness, we convert the R-band magnitudes to flux density in mJy using $F = 10^{(6.489-0.4\times{\rm mag})}$, as suggested by \cite{Mead1990}. For both flux-density and $\Pi$ we estimate the amplitude of the variations by subtracting the value of the lowest BB over the entire dataset, and use the width of the BBs as a proxy for the variability timescale.

Figures \ref{plt:BB_mag} and \ref{plt:BB_pol} show the BB analysis for the flux density and $\Pi$, respectively. For BL Lac, we find similar variability timescales, on average, for both flux-density  and $\Pi$. The average timescale is about 4 h  with a minimum and maximum of 0.9 and $\sim24$ h respectively. $\Pi$ shows more peaks with amplitudes that vary between 3.3\% and 10.6\% with an average of 5\%. The brightness shows four peaks of similar amplitude that range from 13.2 to 19.3~mJy. For both brightness and $\Pi$ the average flaring period is 24 and 25.5 h with a minimum of 16.6 and 18.5 h respectively.  These are comparable to the time scales for optical and X-rays (15 and 14.5 h respectively) found in \cite{Weaver2020}.

For J0211 there is a single peak in brightness and two low amplitude flares in $\Pi$. There is a clear rise in both at the end of our campaign, for which we unfortunately were not able to capture the peak. The amplitude of the peak in brightness is 0.57 mJy and in $\Pi$ is 2.5\%. The average variability timescale is 11.7 h for brightness and 13.3 h for $\Pi$. The minimum and maximum are 3.5 and 23.8 h for brightness and 3 and 34.2 h for the degree of polarization respectively. The duration of the flaring period for the brightness is 75.6 h and for $\Pi$ 31.6 and 55.7 h respectively. Table \ref{tab:BB} summarizes the results from the BB analysis.

\section{Testing particle acceleration models} \label{sec:test_models}

The temporal behavior of the flux and polarization of both BL Lac and J0211+1051 gives the impression that stochastic processes play an important role in driving the variations. This is especially apparent in BL Lac, in which there is $\sim1$ reversal per day in the time derivative of the flux, and $\sim2$ reversals per day in both polarization degree and angle. To understand the origin of such short-term variability, we employ two distinct numerical models that involve stochastic physical processes: magnetic reconnection modeled by particle-in-cell (PIC) simulations, and a scenario in which turbulent plasma crosses a shock. We also test a scenario  were both processes contribute equally to the emission. In this section we present the outcomes of numerical computations based on these models, and contrast the results with the observational data. J0211 exhibits much smoother behavior with only a small number of features. Therefore,  we concentrate on BL Lac, since the NOPE observations sample a large number of reversals of the time derivatives of these observed quantities, and thus provide sufficient statistics to test the ability of the models to reproduce the essence of the observed behavior.

\subsection{PIC magnetic reconnection simulations}

In our realization of the magnetic reconnection model, we assume a pre-existing current sheet moving in the jet direction $z$ with a bulk Lorentz factor $\Gamma=10$ as in \citet{Zhang2020}. The line of sight is perpendicular to the jet propagation direction along the $y$ axis in the comoving frame, so that in the observer's frame the viewing angle of the jet axis is $1/\Gamma$ and the Doppler factor $\delta= \Gamma$. Given that a blazar can have a significant proton population \cite[e.g.,][]{Hovatta2019}, we assume a proton-electron plasma with an initial electron magnetization factor of $\sigma_{\rm e}=4\times 10^4$, which is related to the total magnetization factor by $\sigma_{\rm e}\sim 1836\sigma$, and a guide field $B_{\rm g}=0.2B_0$, where $B_0$ is the anti-parallel component of the magnetic field. Particles initially follow a Maxwell–J\"uttner distribution with the same upstream temperature for electrons and protons, $T_{\rm e}=T_{\rm p}=100m_{\rm e}c^2/k_{\rm B}$, where $k_{\rm B}$ is the Boltzmann constant. We add a radiative reaction force to mimic the effect of cooling, parameterized by $C_{10^4}$. This parameter describes the strength of the radiative cooling at an electron energy of $\gamma_{\rm e}=10^4$, and a smaller value means stronger cooling. Our simulation grid size is $16384\times8192$ in the $x$-$z$ plane, with a physical size of $2L\times L$, where $L=32000 d_{\rm e0}$ and $d_{\rm e0}$ is the non-relativistic electron inertial length. Each cell has 100 particles. The simulation is performed with the \texttt{VPIC} code developed by \citet{Bowers2008}. We output 1600 PIC snapshots and feed into the \texttt{3DPol} code for post-processing the radiation and polarization signatures. In this way, we obtain an optical light curve and polarization with very high time resolution.

Once we have a simulated light curve, we proceed with adding observational noise. This is achieved by estimating the average fractional uncertainty, $\langle$uncertainty/ value$\rangle$, from the observed data, and appropriately assigning an uncertainty value to the simulated data points \citep{Jormanainen2023}. We then replace the simulated data by randomly sampling from a Gaussian distribution, using the value and its uncertainty as the mean and standard deviation of the distribution. Since the time step of the simulations is not related to physical time, we assume that each simulation step is equal to 30, 60, 90, 120, 150, 180 min, and repeat the analysis for each case. For each simulation in each case, we have selected 20 sub-sets from the original simulation and resampled each to the sampling of the observed data by linear interpolation. We caution the reader that, depending on the chosen time interval, subsets may overlap substantially and do not provide fully independent tests. An example of the ``observed'' simulated light curves is shown in Fig. \ref{plt:pic_sims}.

For each new simulation we identify the following properties. We estimate (1) the distribution of polarization degree values based on the Bayesian blocks. We identify the flares in the polarization degree using Bayesian blocks (as above) and calculate the duration and amplitude of each flare, which gives us distributions of (2) the $\Pi$-flare durations and (3) the $\Pi$-flare amplitudes. Furthermore, we identify polarization angle inversions as defined in \citet{Kiehlmann2016} using {\it polarizationtools}\footnote{https://github.com/skiehl/polarizationtools \citep{Kiehlmann2024}}. For each epoch that the polarization angle rotates consistently in one direction we calculate the absolute amplitude, the duration, and the absolute rate, the latter defined as absolute amplitude/duration. This gives us distributions of the polarization angle (4) absolute amplitudes, (5) durations, and (6) absolute rates. We estimated the same distributions from the observed data and use the Anderson-Darling (AD, \citealp{AD1952,AD1954}) test to compare observations with simulations.  The AD test is a non-parametric test with the null hypothesis that two samples come from the same parent distribution, similar to the Kolmogorov-Smirnov test. Figure \ref{plt:pic_hist} shows the observed and simulated property distributions for one example simulation. We use a $5\%$-significance level and count a simulation as successful in reproducing an observed property distribution when the AD-test p-value falls below that threshold.
We repeat these tests for all simulation sub-sets to determine a success rate for each of the six properties separately.

We find that the AD test almost always rejects the $\Pi$ distribution. This is expected, since the simulation considers only a ``naked'' reconnection layer, which naturally produces higher degree of polarization. In reality, this reconnection layer is embedded in a larger-scale jet, possibly acting simultaneously with other reconnection layers and/or other emission regions. Given the endless possibilities and our currently inability to produce global jet PIC magnetic reconnection simulations, we do not attempt to address this issue, and instead focus on normalized quantities such as the $\Pi$ flare amplitudes instead of, e.g., the absolute $\Pi$ flare peak value. A time step of 120 min seems to reproduce the observed light curves best across metrics. Figure \ref{plt:pic_hist} shows the observed and simulated histograms of one realization of the PIC light curves with a time step of 120 min.  On the other hand, for a time step of 30 min  almost all of the simulated light curves cannot reproduce any of the observed behavior. For the remaining time steps, the simulated light curves can reproduce $\Pi$ flares duration and $\Psi$ inversions duration fairly well, however almost all $\Pi$ flare amplitudes and the majority of the absolute $\psi$ amplitude and absolute rate distributions are rejected for all timescales. Overall we find an average success rate across all metrics to vary from 0\% to 97\% (Table \ref{tab:sim_comp}). While the simulations can produce some metrics well, we conclude that they cannot fully reproduce the observed behavior. However, given the caveats and limitations of the current simulation setup capabilities, the high level of agreement with some of the observed properties motivates future work. 

\subsection{Turbulent plasma crossing a shock}

We employ the Turbulent Extreme Multi-Zone (TEMZ) numerical model \citep{Marscher2014,Marscher2021,Marscher2022} in an effort to reproduce the general temporal behavior of the flux and polarization. The model assumes that the variable emission is dominated by synchrotron radiation and synchrotron self-Compton (SSC) scattering \citep{Jones1974} by relativistic electrons (and positrons, if present) that are accelerated as the jet flow crosses a conical standing shock. Any ambient emission outside the shocked region is neglected. The magnetic field of the upstream flow is turbulent, which is realized by dividing the jet into thousands of cells. Each of the cells is approximated to have a unique, uniform magnetic field and density. A given cell belongs to four nested zones of size $1\times1$, $2\times2$, $4\times4$, and $8\times8$ cells$^2$. The magnitude of the magnetic field and density, as well as the direction of the field, are selected at random in each zone from a log-normal distribution. The field and density of a cell are then determined by the average of the values in its zones, weighted according to the Kolmogorov spectrum. In order to approximate the rotating motion of turbulent vortices \citep{Calafut2015}, a turbulent circular 4-velocity about the center of each zone is relativistically added to the systemic flow velocity. 

The TEMZ model assumes that electrons in the jet are already relativistic before crossing the shock, presumably by second-order Fermi acceleration and many non-explosive magnetic reconnection events in the turbulent plasma. The shock then increases the electron energies by a factor of order the Lorentz factor of the flow, although higher when the shock is ``subluminal,'' with magnetic field direction close to the shock normal in the co-moving plasma frame. Although we emphasize here the optical synchrotron radiation, the TEMZ code also calculates the SSC emission, which is important for the energy losses of the electrons. The seed photon field incident on a given cell is calculated by summing the retarded-time contribution from all of the other cells, appropriately Doppler shifted according to their velocities relative to the scattering cell.

The TEMZ simulations are then treated in the same way as the PIC simulations in the section above (see example Fig. \ref{plt:temz_sims}, \ref{plt:temz_hist}). In all the tests the TEMZ simulations can reproduce the $\Pi$ flare amplitudes well and the majority of tests also reproduces the $\Pi$ flare durations well. They are also able to reproduce the $\psi$ inversion duration well with a 60\% success rate (Table \ref{tab:sim_comp}), however, they do not reproduce the $\Pi$ distribution, as well as the absolute $\psi$ amplitude and rate distributions. Similar to the PIC simulations,  we conclude that the TEMZ simulations presented here do not fully reproduce the observed behavior. In this case, we see that some metrics (e.g., $\Pi$ flare durations) can be reproduced with 100\% success rate, while other (e.g., $\Pi$ distribution) have a 0\% success rate. This can be attributed to the vast parameter space of the TEMZ model that cannot be fully explored in a realistic scenario as well as the sensitivity to initial conditions. A large study exploring the TEMZ parameter space as well as a statistical comparison with a larger sample of sources is certainly warranted and can produce a more representative picture. We defer that to future work. 

\subsection{PIC+TEMZ simulations test}

Here we test a scenario with both processes having an equal contribution to the total flux. Since each set of simulations has a different scale for Stokes $I$, before adding them we normalize the simulated light curves in three ways. In the first case, we normalize both light curves such that the minimum is 0 and the maximum is 1. In the second case, we divide the TEMZ simulation by its maximum value. The normalized maximum is then 1, the minimum is $>$0, and the ratio min/max is conserved. The PIC simulations are normalized so that the  minimum equals that of the TEMZ curve and the maximum equals to 1.  In the third case, we normalize TEMZ as in case~2 and PIC as in case~1. In this scenario TEMZ is always present (>0) and the PIC event eventually fades out completely. Except for the strong PIC flares in the beginning, TEMZ is more dominant on average.
We produce simulated light curves for each of the three scenarios with all possible combinations of the TEMZ and PIC model sub-sets. For the PIC model we assume a time sampling of 120 min, which gave the best results for the stand-alone PIC tests above.
We then resample the light curves and for each time-bin we calculate the joined Stokes parameters as $I_{\rm joined} = 0.5 \times I_{\rm TEMZ} + 0.5 \times I_{\rm PIC}$, $ Q_{\rm joined} = 0.5 \times Q_{\rm TEMZ} + 0.5 \times Q_{\rm PIC}$, and $ U_{\rm joined} = 0.5 \times U_{\rm TEMZ} + 0.5 \times U_{\rm PIC}$. Finally, we convert the Stokes parameters to $\Pi$ and $\psi$, apply observational noise, and treat the combined light curve as in the previous cases.  Examples of the differently normalized light curves are shown in the Appendix \ref{app:comp_models}.

We find that the combined light curves can reproduce the $\Pi$ flare duration, however, in almost all tests they were unable to reproduce any of the other metrics (31\% success rate in the best case scenario). Overall, the equal combination of PIC+TEMZ performed poorer than either PIC or TEMZ individually.

\section{Discussion and Conclusions}\label{sec:discussion}

We reported on the first NOPE campaign on BL Lac and J0211 during and outburst and quiescence respectively. The combination of telescopes in different time-zones allowed us to monitor the brightness and polarization variations of both sources for more than the typical eight hour observing nights. Our efforts revealed flaring and interesting behavior that would not have been possible to observe otherwise. 

BL Lac shows rapid, large amplitude, flaring in both brightness and polarization with no signs of correlation between the two. The polarization angle varies rapidly as well, but remains within 10\degr--30\degr\ from the projection of the jet axis. The outburst of BL Lac in 2021 has already been studied in different times and longer timescales \citep{Jorstad2022,Raiteri2023,Imazawa2023}. We find similar behavior regarding the brightness and polarization variability and flare amplitudes, suggesting the underlying mechanism did not change over the course of the outburst. There are two notable differences with previous studies.  \cite{Raiteri2023} reported an anti-correlation between brightness and $\Pi$ and \cite{Jorstad2022} reported quasi-periodic variability. The analysis of \cite{Raiteri2023} includes the time period studied here, while the analysis of \cite{Jorstad2022} stops in August 2021, i.e., two months before our observations. We find none of the aforementioned behavior, however, that could be due to the much shorter time-span of our observations or simply because the quasi-periodic behavior stopped before our campaign. 

On the other hand, J0211 shows smooth variations with a much more complex relation between brightness and polarization including both a correlation and an anti-correlation. During our campaign, we observe a smooth monotonic rotation of the polarization plane of approximately 185$^\circ$. J0211 is a relatively under-studied source that only recently started to attract attention due to its possible association with a high-energy neutrino \citep{Hovatta2021} and potential as an X-ray polarization target \citep{Liodakis2019-II,Peirson2022}. There are only two polarization studies focusing on a flare that took place in 2011  \citep{Chandra2012,Chandra2014}. In the flaring state,  J0211 shows intra-night variability, high $\Pi>20\%$, and large amplitude erratic changes from night-to-night in both polarization degree and angle. This behavior is fairly different from what we observe in quiescence with low $\Pi$ and smooth variations. However, it is very much consistent with the behavior found in BL Lac, suggesting that outbursts in different blazars could be driven by a common mechanism. We can decompose the observed polarization behavior into two polarized components following \cite{Morozova2014}. We split the data in two parts according to the Bayesian block analysis (MJD~59525).  The first part includes the rotation, while the second part the stable $\psi$ and correlation of the brightness and $\Pi$. We find that the two components show similar polarization degree ($\sim9\%$) with slight different $\psi$ ($\sim9\degr$ and $\sim32\degr$ respectively). The flux density of the first component only shows mild variations and is likely associated with the underlying jet emission, while the second, more variable, component shows a 40\% increase in its flux density and could be associated with the emergence of a new shock.  

Such a well-sampled rotation also provides the unique opportunity to test how sampling can affect the detection of such events. We test the effects of cadence by resampling the light curve in different time bins and apply the PA adjustment when necessary. We then check if the resulting light curve is consistent with the observed rotation. We find that for time-bins of 3 h we can always recover the original rotation. However, for a time-bin of 12 h our success rate of recovering the observed rotation is 27\%, and for 1 day none of the resampled light curves show the rotation. This would suggest that large, fast rotations within 1-2 days such as the ones found by IXPE \citep{DiGesu2023} should be more common in the optical for low/intermediate-peaked sources such as BL Lac and J0211.

We compared the NOPE observations to state of the art PIC magnetic reconnection and turbulent plasma simulations focusing on BL Lac. We find that the simulations can adequately reproduce some of the observed polarization properties, but not all of the observed behavior. We attempted to combine the two process in a single simulated light curve, however, the resulting simulations performed worse than the individual processes. We note that there are several caveats in this comparison. Firstly, both the PIC reconnection simulation and TEMZ simulation target generic blazars, i.e., they are not tuned to the parameter space of BL Lac. Since the simulations only sample a small range of parameter space, it is likely that the parameters used in the simulations do not closely match the BL Lac physical conditions of the observed event here. Secondly, simulations generally represent a patch of the blazar zone that is actively accelerating particles, i.e., the flaring region, but do not necessarily consider the emission that comes from other parts of the blazar zone, which is likely the physical origin of the ``quiescent'' blazar emission. This is particularly true for PIC simulations, since the simulations start with a thermal distribution of particles. The quiescent emission typically has low polarization based on previous optical polarization observations, which can change the overall $\Pi$ distribution that is not well described by PIC simulations in our comparison. Thirdly, simulations may not match the physical time of the observed event. This can significantly alter the polarization variability time scales and rate that are not well described by the models in our comparison. Future comparisons of theoretical models and observations may normalize the simulated light curves to match the properties of observed light curves to better constrain the time scales, hence better compare the polarization variability. Finally, multiple particle acceleration mechanisms may occur in the blazar zone. Although we attempted to combine the reconnection and turbulence simulations to compare with observations, our combination of the two models is quite preliminary and oversimplified. Nevertheless, the fact that we already see agreement between observations and simulations is extremely encouraging and motivates both the further development of simulation frameworks as well as similar observing campaigns.

\begin{acknowledgements}
We thank the anonymous referee for useful comments that helped improve this work. The IAA-CSIC co-authors acknowledge financial support from the Spanish "Ministerio de Ciencia e Innovaci\'{o}n" (MCIN/AEI/ 10.13039/501100011033) through the Center of Excellence Severo Ochoa award for the Instituto de Astrof\'{i}isica de Andaluc\'{i}a-CSIC (CEX2021-001131-S), and through grants PID2019-107847RB-C44 and PID2022-139117NB-C44. Some of the data are based on observations collected at the Centro Astron\'{o}mico Hispano en Andalucía (CAHA), operated jointly by Junta de Andaluc\'{i}a and Consejo Superior de Investigaciones Cient\'{i}ficas (IAA-CSIC). E. L. was supported by Academy of Finland projects 317636 and 320045. The data in this study include observations made with the Nordic Optical Telescope, owned in collaboration by the University of Turku and Aarhus University, and operated jointly by Aarhus University, the University of Turku and the University of Oslo, representing Denmark, Finland and Norway, the University of Iceland and Stockholm University at the Observatorio del Roque de los Muchachos, La Palma, Spain, of the Instituto de Astrofisica de Canarias. The data presented here were obtained in part with ALFOSC, which is provided by the Instituto de Astrof\'{\i}sica de Andaluc\'{\i}a (IAA) under a joint agreement with the University of Copenhagen and NOT. We acknowledge funding to support our NOT observations from the Finnish Centre for Astronomy with ESO (FINCA), University of Turku, Finland (Academy of Finland grant nr 306531). This research has made use of data from the RoboPol program, a collaboration between Caltech, the University of Crete, IA-FORTH, IUCAA, the MPIfR, and the Nicolaus Copernicus University, which was conducted at Skinakas Observatory in Crete, Greece. D.B. and S.K. acknowledge support from the European Research Council (ERC) under the European Unions Horizon 2020 research and innovation program under grant agreement No.~771282. JS and SR acknowledge the partial support of SERB-DST, New Delhi, through SRG grant no. SRG/2021/001334. NS acknowledges the financial support provided under the National Post-doctoral Fellowship (NPDF; Sanction Number: PDF/2022/001040) by the Science \& Engineering Research Board (SERB), a statutory body of the Department of Science \& Technology (DST), Government of India. KW gratefully acknowledges support from Royal Society Research Grant RG170230 (PI R. Starling), which funded the LE2Pol polarimeter. The Liverpool Telescope is operated on the island of La Palma by Liverpool John Moores University in the Spanish Observatorio del Roque de los Muchachos of the Instituto de Astrofisica de Canarias with financial support from the UK Science and Technology Facilities Council. R.I. is supported by JST, the establishment of university fellowships towards the creation of science technology innovation, Grant Number JPMJFS2129. I.L was supported by the NASA Postdoctoral Program at the Marshall Space Flight Center, administered by Oak Ridge Associated Universities under contract with NASA. This work was supported by JSPS KAKENHI Grant Number 21H01137. I.L and S.K. were funded by the European Union ERC-2022-STG - BOOTES - 101076343. Views and opinions expressed are however those of the author(s) only and do not necessarily reflect those of the European Union or the European Research Council Executive Agency. Neither the European Union nor the granting authority can be held responsible for them. The effort at Boston University was supported in part by National Science Foundation grant AST-2108622 and and NASA Fermi Guest Investigator grants 80NSSC22K1571 and 80NSSC23K1507. This study used observations conducted with the 1.8 m Perkins Telescope Observatory (PTO) in Arizona (USA), which is owned and operated by Boston University. This work is partly based upon observations carried out at the Observatorio Astronómico Nacional on the Sierra San Pedro Mártir (OAN-SPM), Baja California, México. C.C. acknowledges support by the European Research Council (ERC) under the HORIZON ERC Grants 2021 programme under grant agreement No. 101040021. E. B. acknowledges support from DGAPA-PAPIIT UNAM through grant IN119123 and from
CONAHCYT Grant CF-2023-G-100. ES acknowledges ANID BASAL project FB210003 and Gemini ANID ASTRO21-0003. Observations with the SAO RAS telescopes are supported by the Ministry of Science and Higher Education of the Russian Federation.
\end{acknowledgements}

\appendix

\section{Comparison to simulations}\label{app:comp_models}
\begin{figure*}
    \centering
    \includegraphics[width=\textwidth]{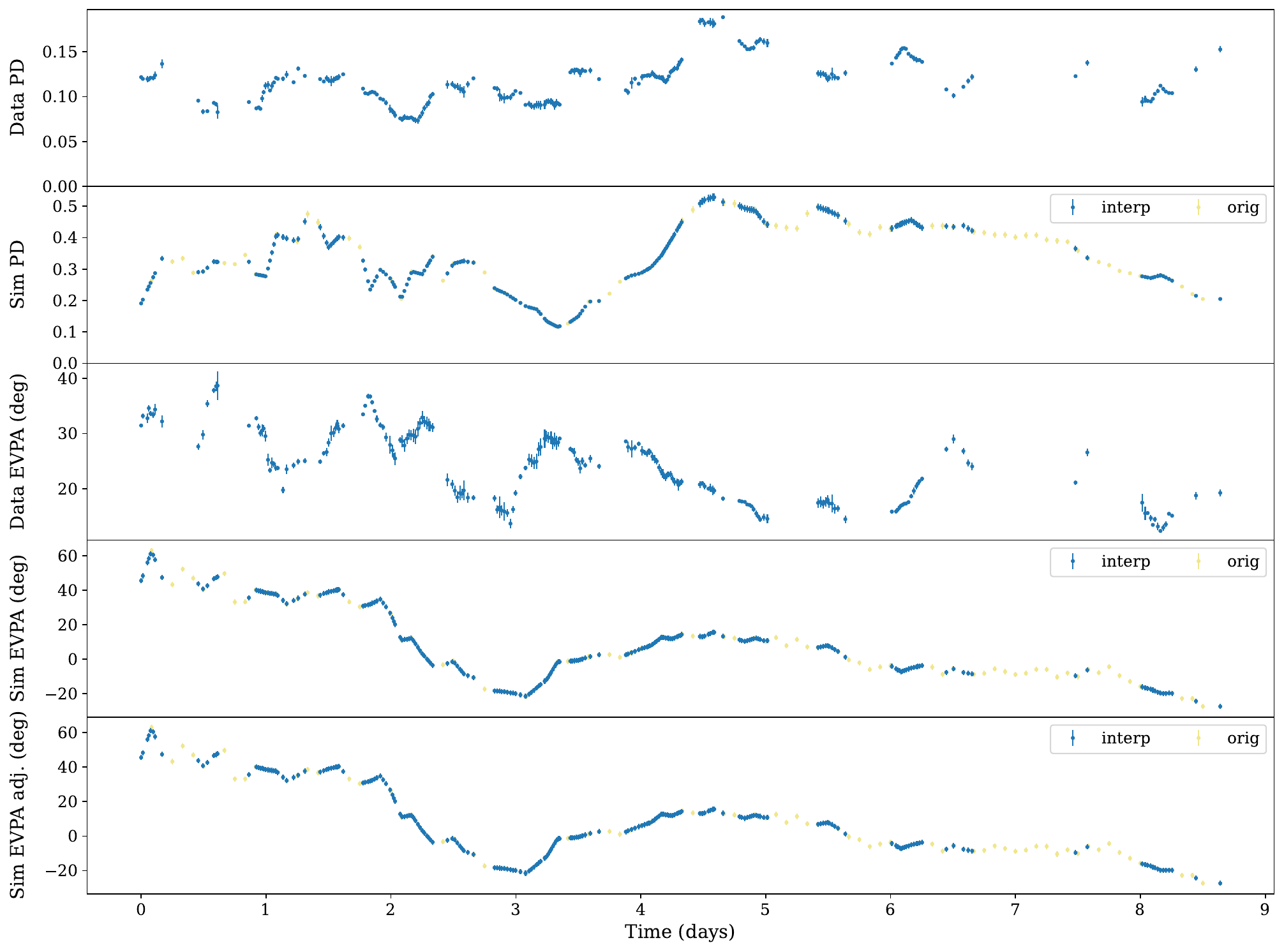}
    \caption{Example light curve for the PIC simulations with a 120\,m time-step. The panels from top to bottom show the observed $\Pi$, the simulated $\Pi$, the observed $\psi$, the simulated $\psi$, and the adjusted simulated $\psi$. The original simulations are shown in yellow, the simulated light curves matching the observed cadence are shown in blue.}
    \label{plt:pic_sims}
\end{figure*}

\begin{figure*}
    \centering
     \includegraphics[width=\textwidth]{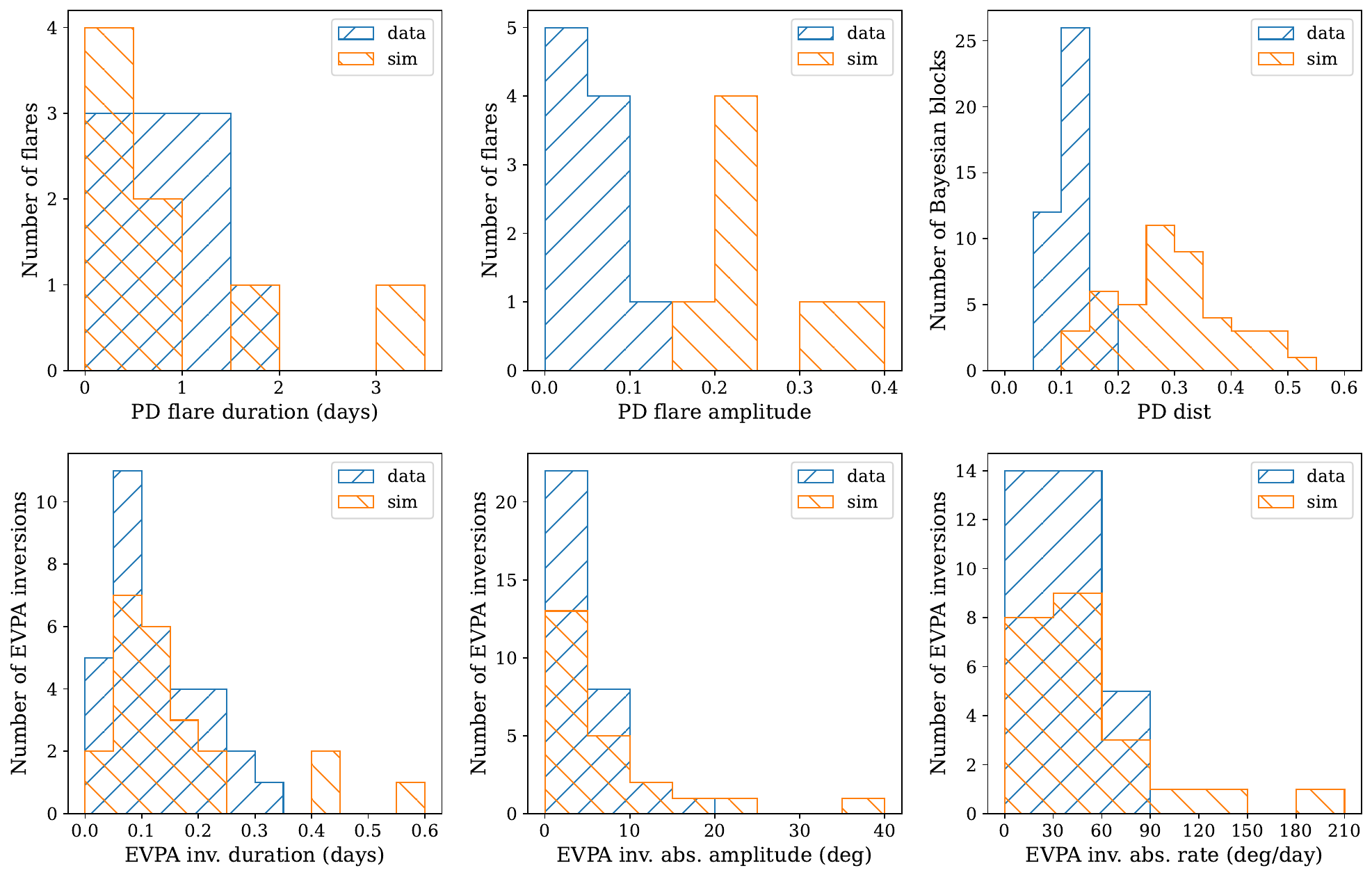}
    \caption{Histograms of the comparison between the observed data and the PIC simulations for the light curves in Fig. \ref{plt:pic_sims}. The top row (left to right) shows $\Pi$ flare duration, $\Pi$ flare amplitude, and $\Pi$ distribution. The bottom row (left to right) shows $\psi$ inversion duration,  $\psi$ absolute inversion amplitude, and $\psi$ absolute inversion rate. The observed data are shown in blue and the simulated in orange.}
    \label{plt:pic_hist}
\end{figure*}

\begin{figure*}
    \centering
    \includegraphics[width=\textwidth]{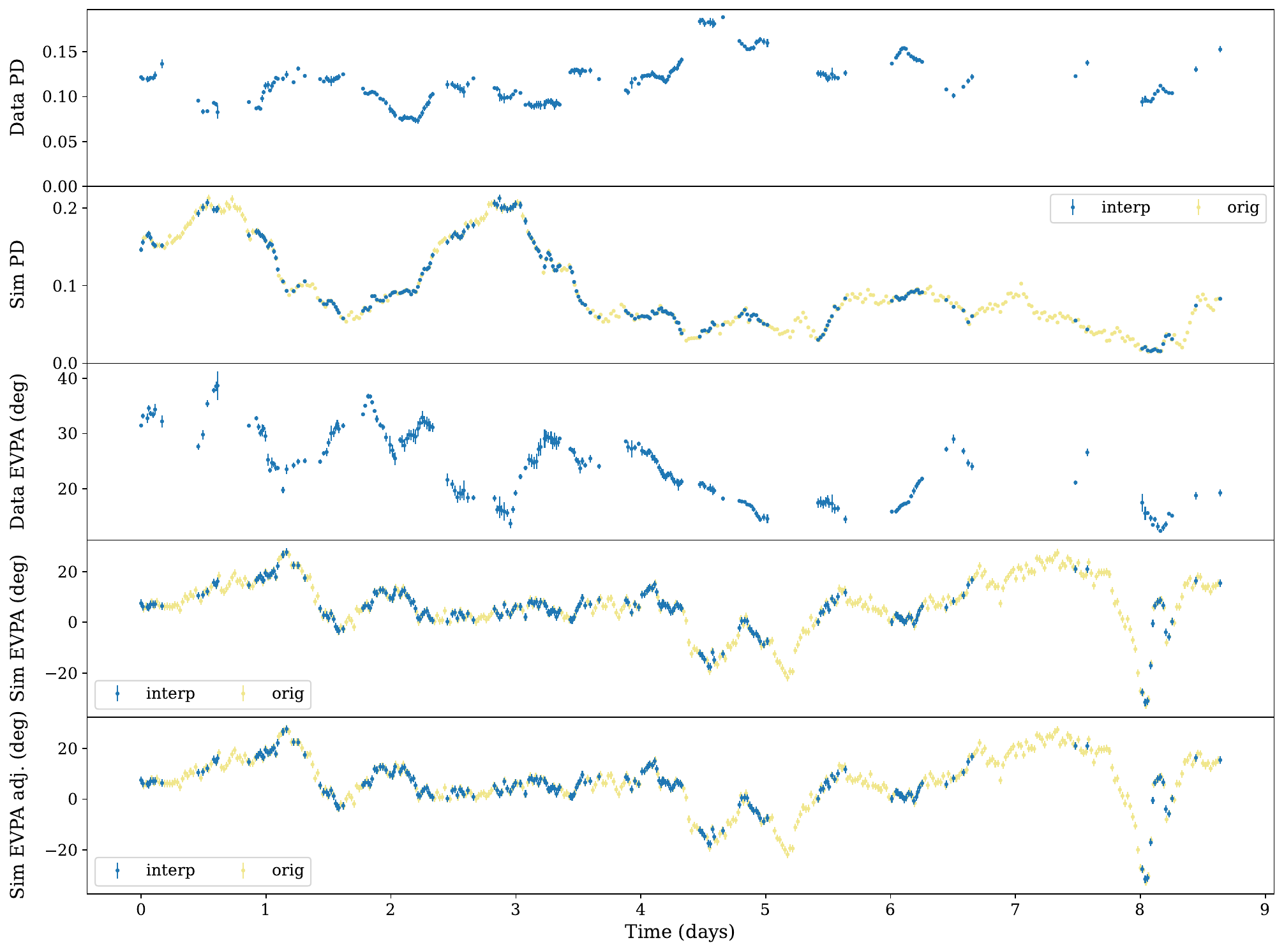}
    \caption{Example light curve for the TEMZ simulations.  Panels as in Fig. \ref{plt:pic_sims}}
    \label{plt:temz_sims}
\end{figure*}

\begin{figure*}
    \centering
     \includegraphics[width=\textwidth]{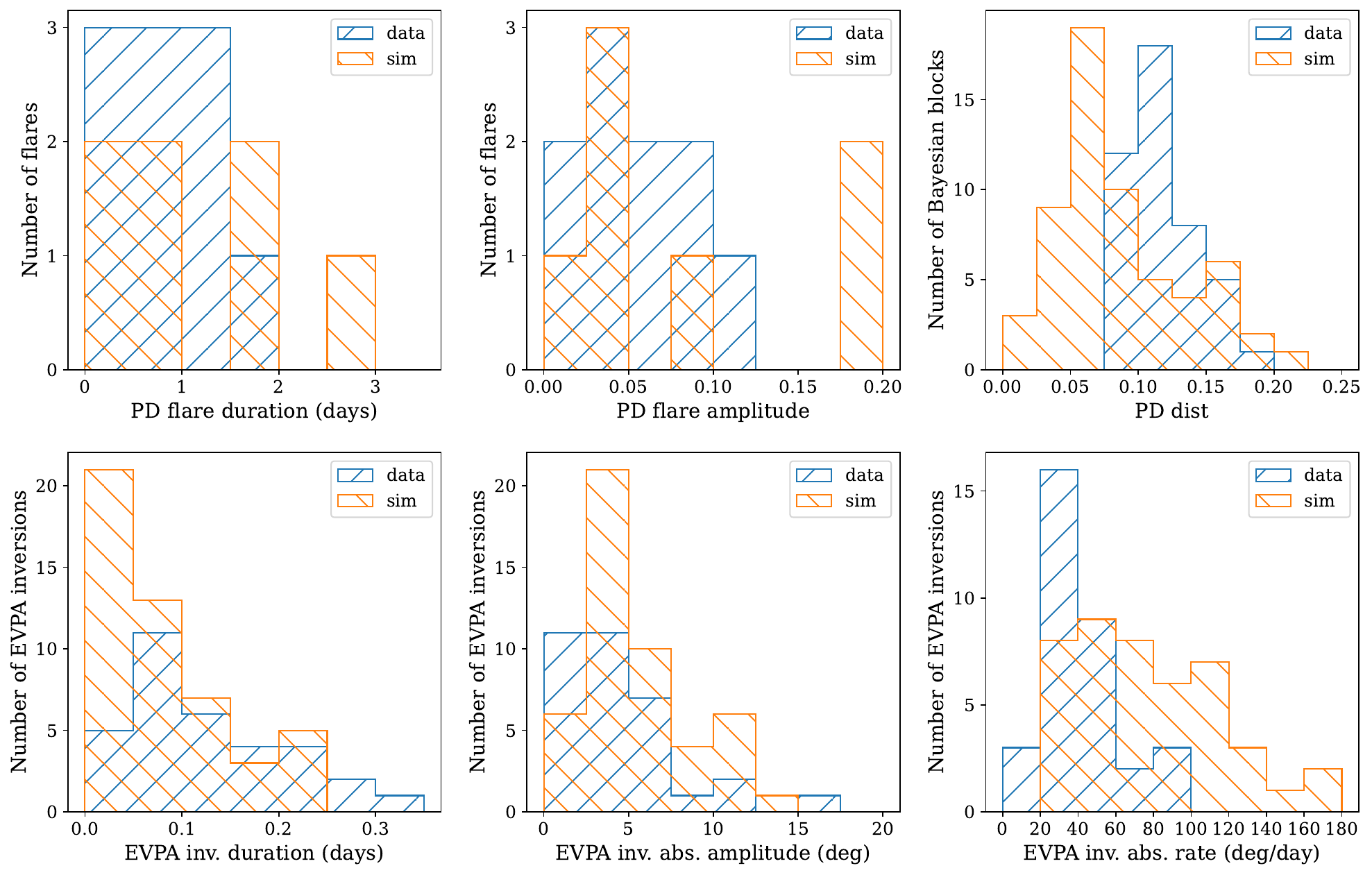}
    \caption{Histograms of the comparison between the observed data and the TEMZ simulations for the light curves in Fig. \ref{plt:temz_sims}. Panels as in Fig. \ref{plt:pic_hist}. }
    \label{plt:temz_hist}
\end{figure*}

\begin{figure*}
    \centering
    \includegraphics[width=\textwidth]{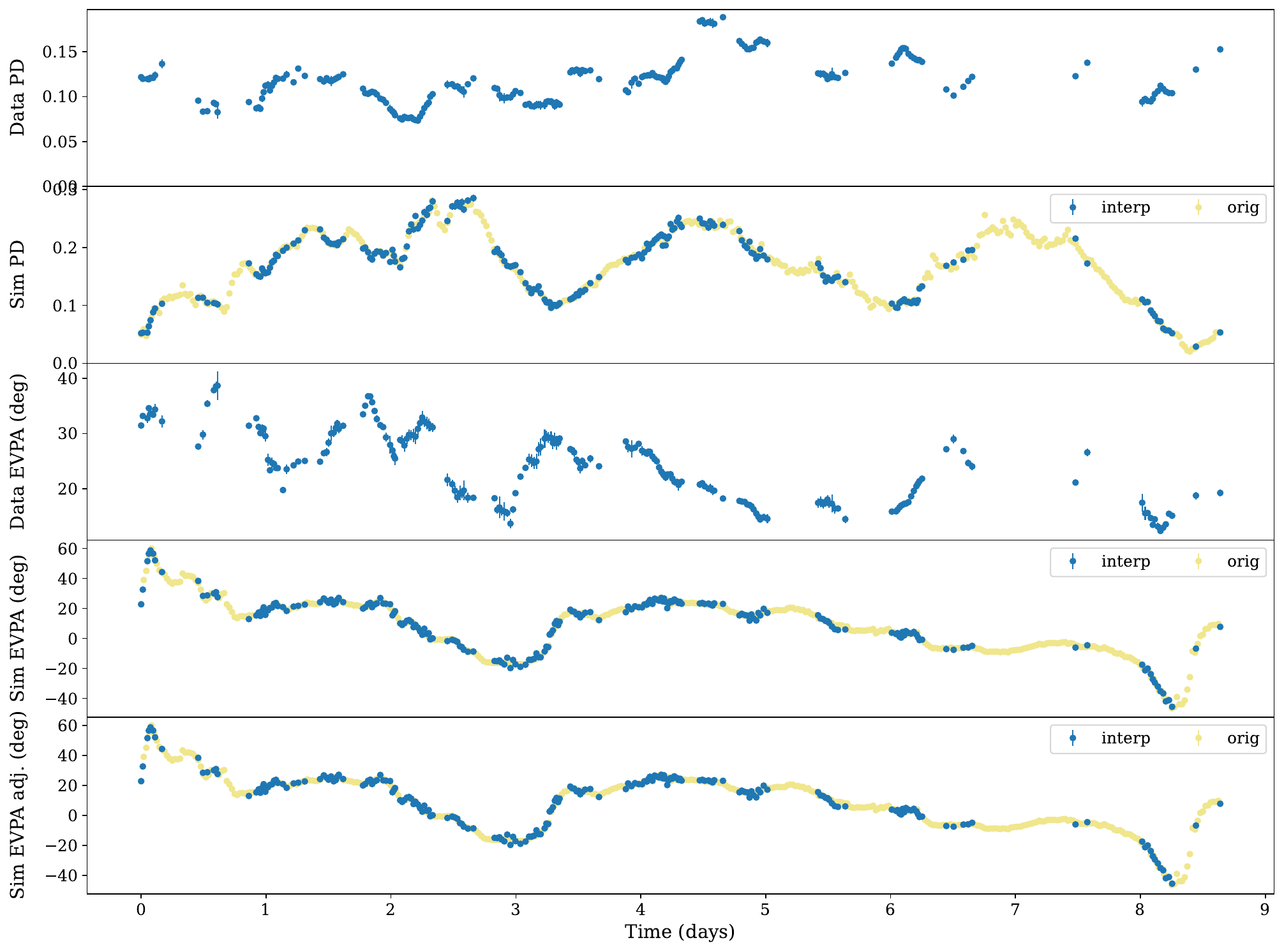}
    \caption{Example light curve for the PIC+TEMZ simulations (Norm 1).  Panels as in Fig. \ref{plt:pic_sims}}
    \label{plt:temzpic_sims}
\end{figure*}

\begin{figure*}
    \centering
     \includegraphics[width=\textwidth]{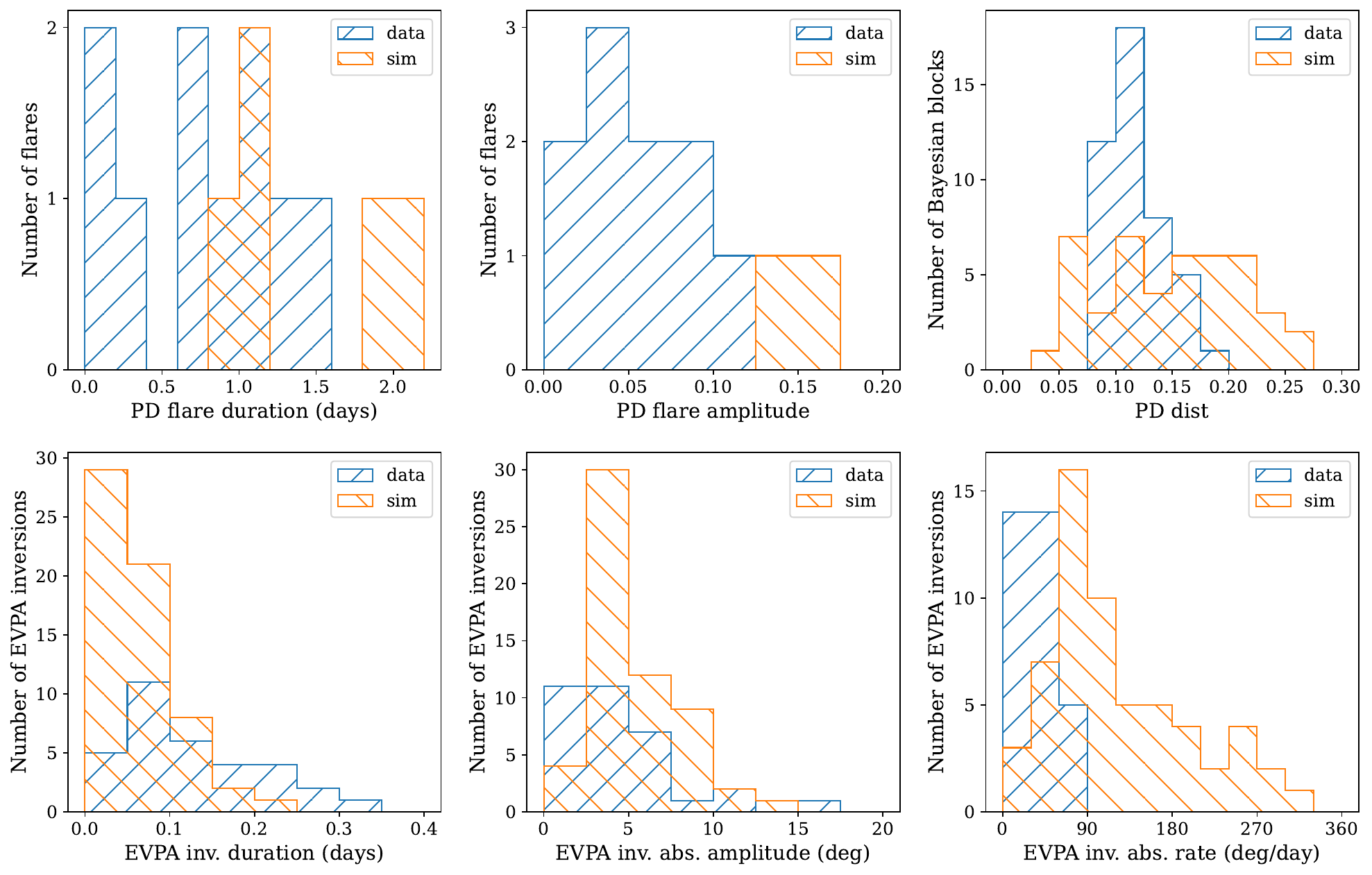}
    \caption{Histograms of the comparison between the observed data and the PIC+TEMZ simulations for the light curves in Fig. \ref{plt:temzpic_sims}. Panels as in Fig. \ref{plt:pic_hist}. }
    \label{plt:temzpic_hist}
\end{figure*}

\begin{figure*}
    \centering
    \includegraphics[width=\textwidth]{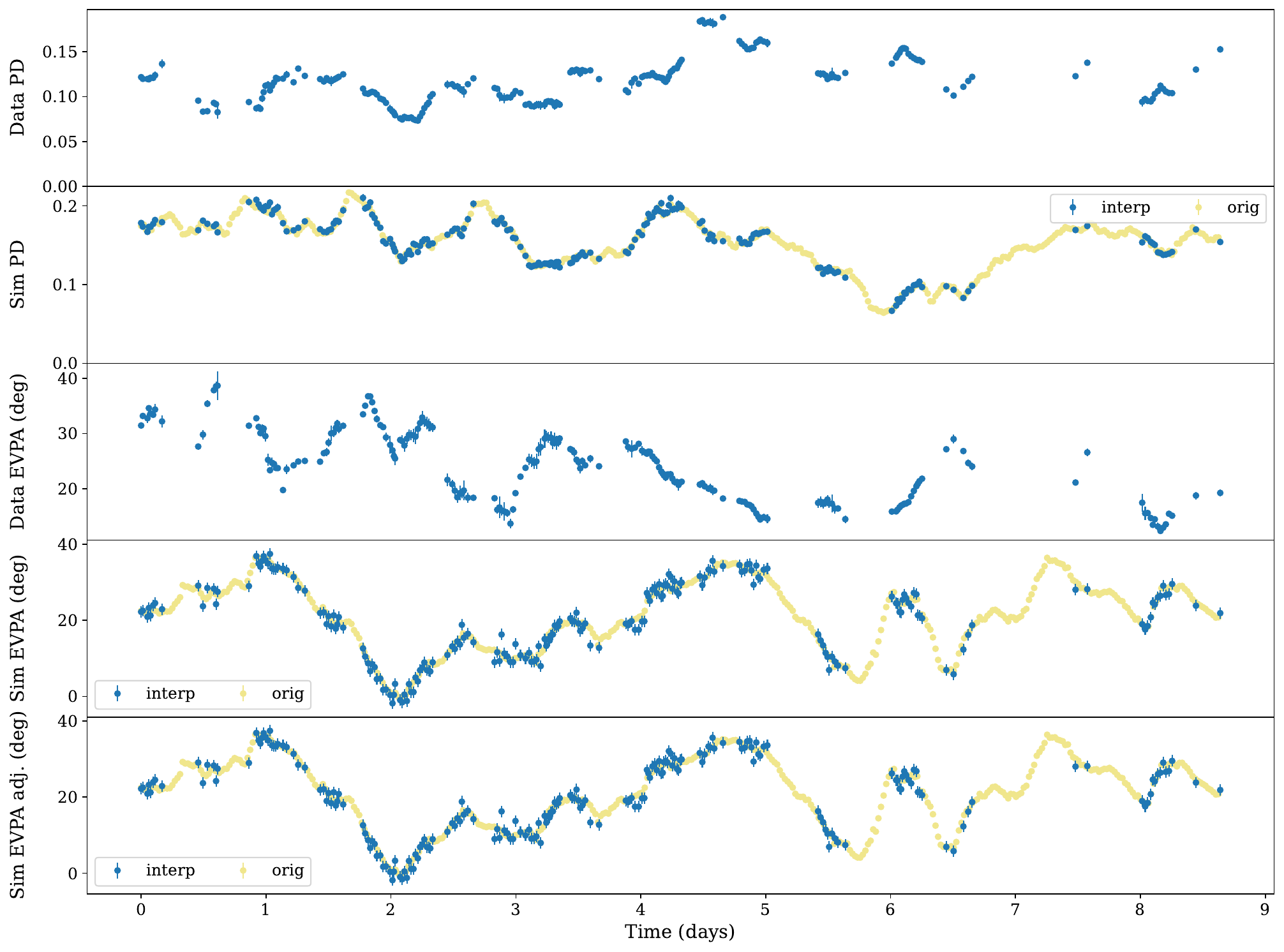}
    \caption{Example light curve for the PIC+TEMZ simulations (Norm 2).  Panels as in Fig. \ref{plt:pic_sims}}
    \label{plt:temzpic_sims2}
\end{figure*}

\begin{figure*}
    \centering
     \includegraphics[width=\textwidth]{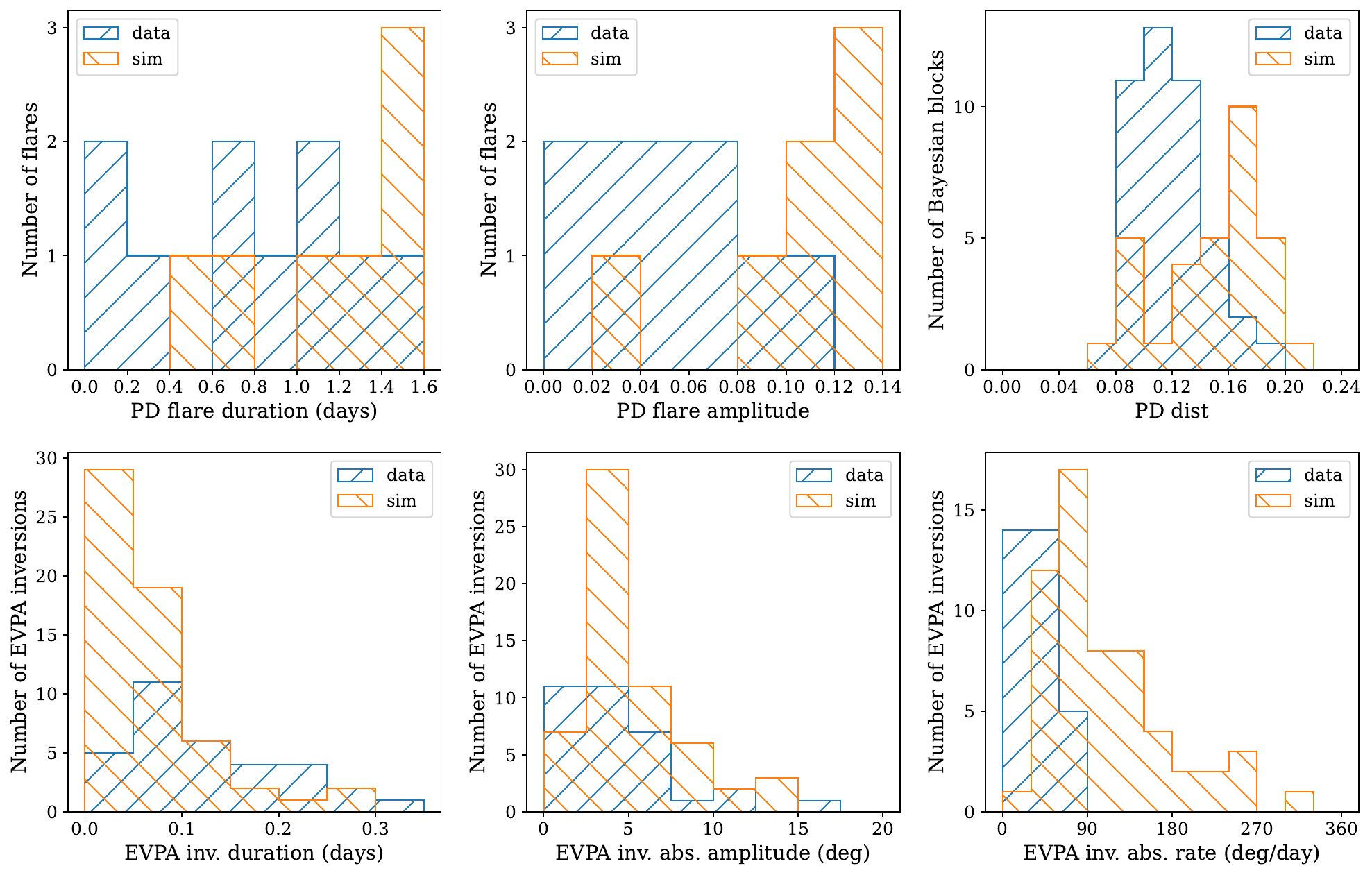}
    \caption{Histograms of the comparison between the observed the PIC+TEMZ simulations for the light curves in Fig. \ref{plt:temzpic_sims2}. Panels as in Fig. \ref{plt:pic_hist}. }
    \label{plt:temzpic_hist2}
\end{figure*}

\begin{figure*}
    \centering
    \includegraphics[width=\textwidth]{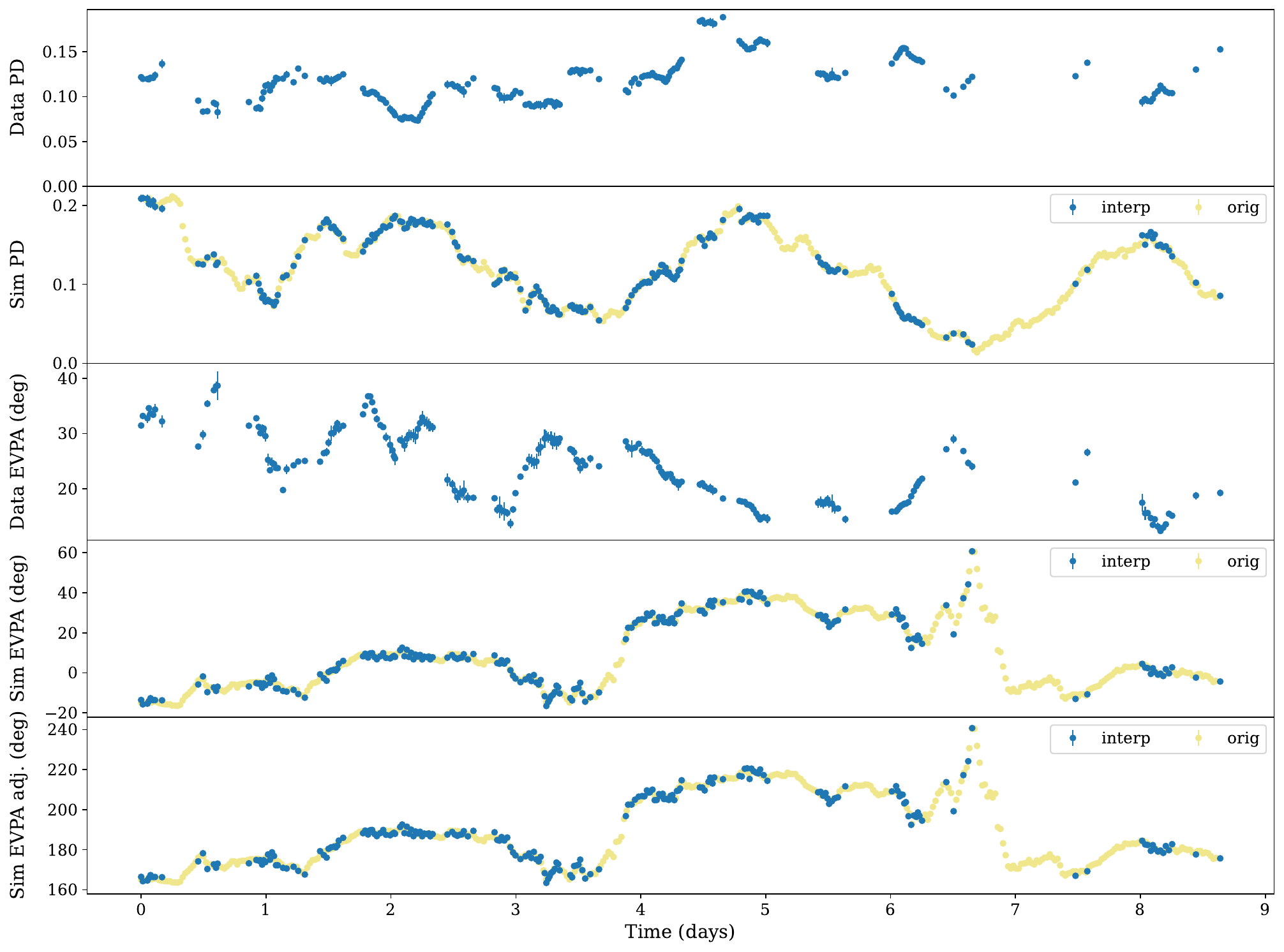}
    \caption{Example light curve for the PIC+TEMZ simulations (Norm 3).  Panels as in Fig. \ref{plt:pic_sims}}
    \label{plt:temzpic_sims3}
\end{figure*}

\begin{figure*}
    \centering
     \includegraphics[width=\textwidth]{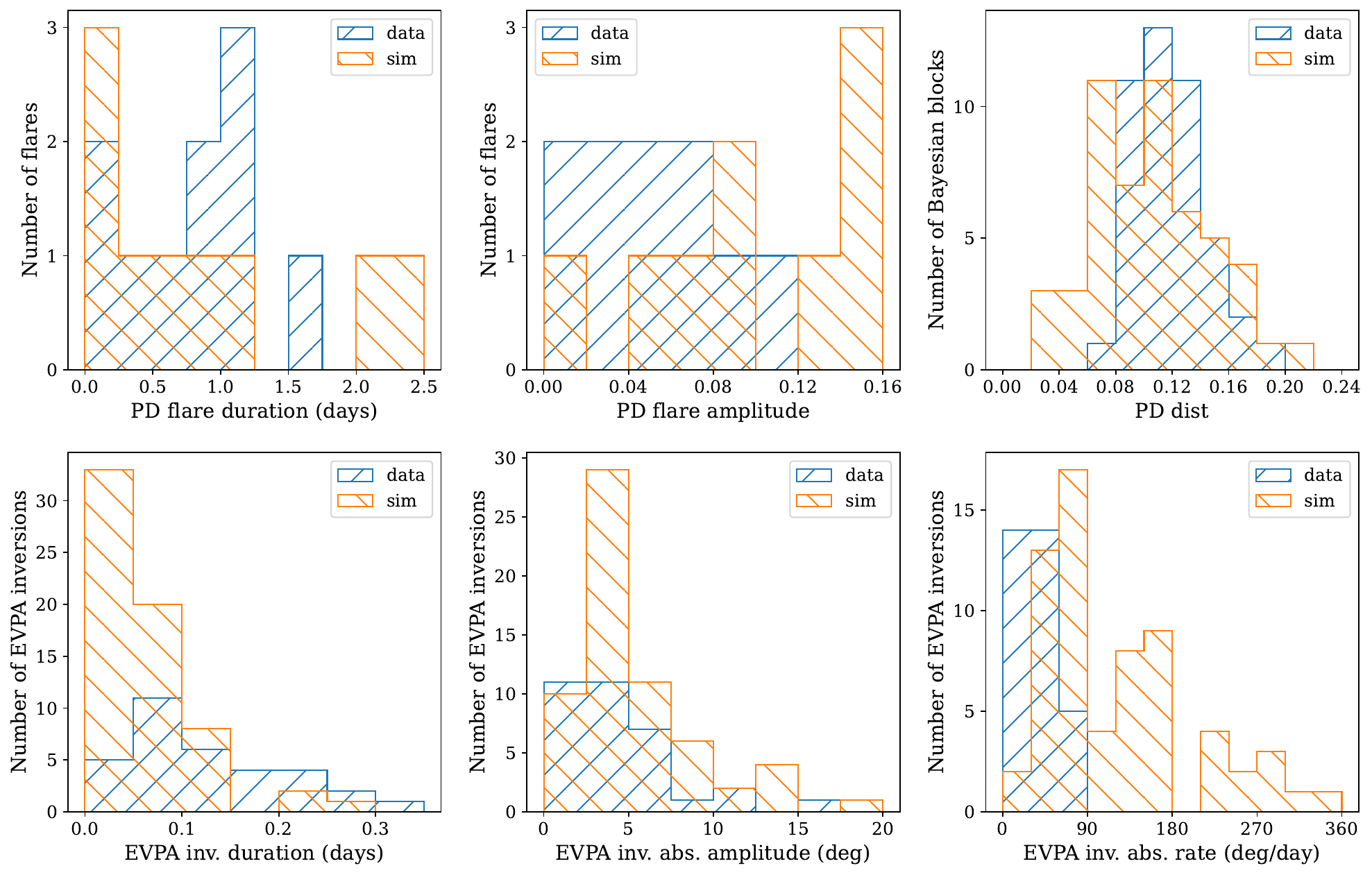}
    \caption{Histograms of the comparison between the observed the PIC+TEMZ simulations for the light curves in Fig. \ref{plt:temzpic_sims3}. Panels as in Fig. \ref{plt:pic_hist}. }
    \label{plt:temzpic_hist3}
\end{figure*}

Table \ref{tab:sim_comp} summarizes the success rate between the observations and the different models discussed in section \ref{sec:test_models}. Figures \ref{plt:pic_sims} and \ref{plt:pic_hist} show examples of the comparison with the PIC simulations, Figs. \ref{plt:temz_sims} and \ref{plt:temz_hist} show examples of the comparison with the TEMZ simulations, and Figs. \ref{plt:temzpic_sims}--\ref{plt:temzpic_sims3}, \ref{plt:temzpic_hist3} examples of the comparison with the TEMZ+PIC simulations for three different normalized simulated light curves (see Table \ref{tab:sim_comp}).


\begin{table*}
\setlength{\tabcolsep}{11pt}
\centering
  \caption{Summary of success rate (\%) between observed and simulated light curves. The columns are (1) $\Pi$ distribution; (2) $\Pi$ flare duration; (3) $\Pi$ flare amplitude; (4) $\psi$ inversion duration; (5) $\psi$ absolute inversion amplitude; (6) $\psi$ absolute inversion rate. For the combined  PIC + TEMZ comparison, we are listing three cases. For the case ``Norm. 1'', both models are normalized between [0,1]. For the case ``Norm 2'', PIC is normalized between [0,1] and TEMZ normalized to the PIC light curve, and for the case ``Norm 3'' TEMZ is normalized between [0,1] and PIC normalized to the TEMZ light curve. For the PIC simulations we have chosen the time step that produces the best agreement with the observations.}
  \label{tab:sim_comp}
\begin{tabular}{@{}lcccccccc@{}}
 \hline
 \hline
Model &$\Pi$  & $\Pi$ &  $\Pi$ & $\psi$ & $\psi$ & $\psi$ \\
& Distr. & Fl. Dur. & Fl. Ampl. & Dur. & Abs. ampl. & Abs. rate\\
  \hline
PIC (120min) & 0 & 97  & 2 & 85 & 8 & 30\\
\hline
TEMZ & 0 & 100 & 55 & 60 & 0 & 0\\
\hline
PIC + TEMZ (norm. 1) & 3 & 98 & 13 & 5 & 0 & 0\\
PIC + TEMZ (norm 2) & 3 & 96  & 31 & 4 & 1 & 0\\
PIC + TEMZ (norm. 3) & 3 & 95 & 9 & 6 & 1 & 0\\
\hline
\end{tabular}
\end{table*}

\bibliographystyle{aa} 

\begin{thebibliography}{93}
\expandafter\ifx\csname natexlab\endcsname\relax\def\natexlab#1{#1}\fi

\bibitem[{{Abdollahi} {et~al.}(2023){Abdollahi}, {Ajello}, {Baldini}, {Ballet},
  {Bastieri}, {Becerra Gonzalez}, {Bellazzini}, {Berretta}, {Bissaldi},
  {Bonino}, {Brill}, {Bruel}, {Burns}, {Buson}, {Cameron}, {Caputo}, {Caraveo},
  {Cibrario}, {Ciprini}, {Cristarella Orestano}, {Crnogorcevic}, {Cutini},
  {D'Ammando}, {De Gaetano}, {Digel}, {Di Lalla}, {Di Venere},
  {Dom{\'\i}nguez}, {Ramazani}, {Fegan}, {Ferrara}, {Fiori}, {Fleischhack},
  {Franckowiak}, {Fukazawa}, {Fusco}, {Gammaldi}, {Gargano}, {Garrappa},
  {Gasbarra}, {Gasparrini}, {Giglietto}, {Giordano}, {Giroletti}, {Green},
  {Grenier}, {Guiriec}, {Gustafsson}, {Hays}, {Horan}, {Hou},
  {J{\'o}hannesson}, {Kerr}, {Kocevski}, {Kuss}, {Latronico}, {Li}, {Liodakis},
  {Longo}, {Loparco}, {Lorusso}, {Lott}, {Lovellette}, {Lubrano}, {Maldera},
  {Manfreda}, {Mart{\'\i}-Devesa}, {Mazziotta}, {Mereu}, {Meyer}, {Michelson},
  {Mizuno}, {Monzani}, {Morselli}, {Moskalenko}, {Negro}, {Omodei}, {Orlando},
  {Ormes}, {Paneque}, {Panzarini}, {Perkins}, {Persic}, {Pesce-Rollins},
  {Pillera}, {Porter}, {Principe}, {Racusin}, {Rain{\`o}}, {Rando}, {Rani},
  {Razzano}, {Razzaque}, {Reimer}, {Reimer}, {S{\'a}nchez-Conde}, {Parkinson},
  {Scargle}, {Scotton}, {Serini}, {Sgr{\`o}}, {Siskind}, {Spandre}, {Spinelli},
  {Suson}, {Tajima}, {Thompson}, {Torres}, {Valverde}, {Venters}, {Wadiasingh},
  {Wagner}, \& {Wood}}]{repository2023}
{Abdollahi}, S., {Ajello}, M., {Baldini}, L., {et~al.} 2023, \apjs, 265, 31

\bibitem[{{Ackermann} {et~al.}(2016){Ackermann}, {Anantua}, {Asano}, {Baldini},
  {Barbiellini}, {Bastieri}, {Becerra Gonzalez}, {Bellazzini}, {Bissaldi},
  {Blandford}, {Bloom}, {Bonino}, {Bottacini}, {Bruel}, {Buehler}, {Caliandro},
  {Cameron}, {Caragiulo}, {Caraveo}, {Cavazzuti}, {Cecchi}, {Cheung}, {Chiang},
  {Chiaro}, {Ciprini}, {Cohen-Tanugi}, {Costanza}, {Cutini}, {D'Ammando}, {de
  Palma}, {Desiante}, {Digel}, {Di Lalla}, {Di Mauro}, {Di Venere}, {Drell},
  {Favuzzi}, {Fegan}, {Ferrara}, {Fukazawa}, {Funk}, {Fusco}, {Gargano},
  {Gasparrini}, {Giglietto}, {Giordano}, {Giroletti}, {Grenier}, {Guillemot},
  {Guiriec}, {Hayashida}, {Hays}, {Horan}, {J{\'o}hannesson}, {Kensei},
  {Kocevski}, {Kuss}, {La Mura}, {Larsson}, {Latronico}, {Li}, {Longo},
  {Loparco}, {Lott}, {Lovellette}, {Lubrano}, {Madejski}, {Magill}, {Maldera},
  {Manfreda}, {Mayer}, {Mazziotta}, {Michelson}, {Mirabal}, {Mizuno},
  {Monzani}, {Morselli}, {Moskalenko}, {Nalewajko}, {Negro}, {Nuss}, {Ohsugi},
  {Orlando}, {Paneque}, {Perkins}, {Pesce-Rollins}, {Piron}, {Pivato},
  {Porter}, {Principe}, {Rando}, {Razzano}, {Razzaque}, {Reimer}, {Scargle},
  {Sgr{\`o}}, {Sikora}, {Simone}, {Siskind}, {Spada}, {Spinelli}, {Stawarz},
  {Thayer}, {Thompson}, {Torres}, {Troja}, {Uchiyama}, {Yuan}, \&
  {Zimmer}}]{Ackermann2016}
{Ackermann}, M., {Anantua}, R., {Asano}, K., {et~al.} 2016, \apjl, 824, L20

\bibitem[{{Afanasiev} {et~al.}(2023){Afanasiev}, {Malygin}, {Shablovinskaya},
  {Uklein}, {Amirkhanyan}, {Perepelitsyn}, \&
  {Afanasieva}}]{2023RASTI...2..657A}
{Afanasiev}, V.~L., {Malygin}, E.~A., {Shablovinskaya}, E.~S., {et~al.} 2023,
  RAS Techniques and Instruments, 2, 657

\bibitem[{{Ajello} {et~al.}(2020){Ajello}, {Angioni}, {Axelsson}, {Ballet},
  {Barbiellini}, {Bastieri}, {Becerra Gonzalez}, {Bellazzini}, {Bissaldi},
  {Bloom}, {Bonino}, {Bottacini}, {Bruel}, {Buson}, {Cafardo}, {Cameron},
  {Cavazzuti}, {Chen}, {Cheung}, {Ciprini}, {Costantin}, {Cutini}, {D'Ammando},
  {de la Torre Luque}, {de Menezes}, {de Palma}, {Desai}, {Di Lalla}, {Di
  Venere}, {Dom{\'\i}nguez}, {Dirirsa}, {Ferrara}, {Finke}, {Franckowiak},
  {Fukazawa}, {Funk}, {Fusco}, {Gargano}, {Garrappa}, {Gasparrini},
  {Giglietto}, {Giordano}, {Giroletti}, {Green}, {Grenier}, {Guiriec},
  {Harita}, {Hays}, {Horan}, {Itoh}, {J{\'o}hannesson}, {Kovac'evic'},
  {Krauss}, {Kreter}, {Kuss}, {Larsson}, {Leto}, {Li}, {Liodakis}, {Longo},
  {Loparco}, {Lott}, {Lovellette}, {Lubrano}, {Madejski}, {Maldera},
  {Manfreda}, {Mart{\'\i}-Devesa}, {Massaro}, {Mazziotta}, {Mereu}, {Meyer},
  {Migliori}, {Mirabal}, {Mizuno}, {Monzani}, {Morselli}, {Moskalenko},
  {Negro}, {Nemmen}, {Nuss}, {Ojha}, {Ojha}, {Omodei}, {Orienti}, {Orlando},
  {Ormes}, {Paliya}, {Pei}, {Pe{\~n}a-Herazo}, {Persic}, {Pesce-Rollins},
  {Petrov}, {Piron}, {Poon}, {Principe}, {Rain{\`o}}, {Rando}, {Rani},
  {Razzano}, {Razzaque}, {Reimer}, {Reimer}, {Schinzel}, {Serini}, {Sgr{\`o}},
  {Siskind}, {Spandre}, {Spinelli}, {Suson}, {Tachibana}, {Thompson}, {Torres},
  {Torresi}, {Troja}, {Valverde}, {van Zyl}, \& {Yassine}}]{Ajello2020}
{Ajello}, M., {Angioni}, R., {Axelsson}, M., {et~al.} 2020, \apj, 892, 105

\bibitem[{Anderson \& Darling(1952)}]{AD1952}
Anderson, T.~W. \& Darling, D.~A. 1952, The Annals of Mathematical Statistics,
  23, 193

\bibitem[{Anderson \& Darling(1954)}]{AD1954}
Anderson, T.~W. \& Darling, D.~A. 1954, Journal of the American Statistical
  Association, 49, 765

\bibitem[{{Baldini} {et~al.}(2021){Baldini}, {Ballet}, {Bastieri}, {Becerra
  Gonzalez}, {Bellazzini}, {Berretta}, {Bissaldi}, {Blandford}, {Bloom},
  {Bonino}, {Bottacini}, {Bruel}, {Buson}, {Cameron}, {Caraveo}, {Cavazzuti},
  {Chen}, {Chiaro}, {Ciangottini}, {Cibario}, {Ciprini}, {Cristarella
  Orestano}, {Crnogorcevic}, {Cutini}, {D'Ammando}, {de la Torre Luque}, {de
  Palma}, {Digel}, {Di Lalla}, {Dirirsa}, {Di Venere}, {Dom{\'\i}nguez},
  {Fiori}, {Fleischhack}, {Franckowiak}, {Fukazawa}, {Funk}, {Fusco},
  {Gargano}, {Gasparrini}, {Germani}, {Giglietto}, {Giordano}, {Giroletti},
  {Green}, {Grenier}, {Griffin}, {Guiriec}, {Gustafsson}, {Hewitt}, {Horan},
  {Imazawa}, {J{\'o}hannesson}, {Kerr}, {Kocevski}, {Kuss}, {Larsson},
  {Latronico}, {Li}, {Liodakis}, {Longo}, {Loparco}, {Lovellette}, {Lubrano},
  {Maldera}, {Manfreda}, {Mart{\'\i}-Devesa}, {Matake}, {Mazziotta}, {Mereu},
  {Meyer}, {Mirabal}, {Mitthumsiri}, {Mizuno}, {Monzani}, {Morselli},
  {Moskalenko}, {Nagasawa}, {Negro}, {Ojha}, {Orienti}, {Orlando},
  {Palatiello}, {Paliya}, {Paneque}, {Pei}, {Persic}, {Pesce-Rollins},
  {Petrosian}, {Poon}, {Porter}, {Principe}, {Racusin}, {Rain{\`o}}, {Rando},
  {Rani}, {Razzano}, {Razzaque}, {Reimer}, {Reimer}, {Saz Parkinson},
  {Scotton}, {Serini}, {Sgr{\`o}}, {Siskind}, {Spandre}, {Spinelli}, {Suson},
  {Tajima}, {Tak}, {Torres}, {Tosti}, {Troja}, {Wood}, {Yassine}, {Zaharijas},
  \& {Fermi-LAT Collaboration}}]{Baldini2021}
{Baldini}, L., {Ballet}, J., {Bastieri}, D., {et~al.} 2021, \apjs, 256, 13

\bibitem[{{Bhatta} {et~al.}(2016){Bhatta}, {Stawarz}, {Ostrowski}, {Markowitz},
  {Akitaya}, {Arkharov}, {Bachev}, {Ben{\'\i}tez}, {Borman}, {Carosati},
  {Cason}, {Chanishvili}, {Damljanovic}, {Dhalla}, {Frasca}, {Hiriart}, {Hu},
  {Itoh}, {Jableka}, {Jorstad}, {Jovanovic}, {Kawabata}, {Klimanov},
  {Kurtanidze}, {Larionov}, {Laurence}, {Leto}, {Marscher}, {Moody},
  {Moritani}, {Ohlert}, {Di Paola}, {Raiteri}, {Rizzi}, {Sadun}, {Sasada},
  {Sergeev}, {Strigachev}, {Takaki}, {Troitsky}, {Ui}, {Villata}, {Vince},
  {Webb}, {Yoshida}, \& {Zola}}]{Bhatta2016}
{Bhatta}, G., {Stawarz}, {\L}., {Ostrowski}, M., {et~al.} 2016, \apj, 831, 92

\bibitem[{{Blandford} {et~al.}(2019){Blandford}, {Meier}, \&
  {Readhead}}]{Blandford2019}
{Blandford}, R., {Meier}, D., \& {Readhead}, A. 2019, \araa, 57, 467

\bibitem[{{Blinov} {et~al.}(2021){Blinov}, {Kiehlmann}, {Pavlidou},
  {Panopoulou}, {Skalidis}, {Angelakis}, {Casadio}, {Einoder}, {Hovatta},
  {Kokolakis}, {Kougentakis}, {Kus}, {Kylafis}, {Kyritsis}, {Lalakos},
  {Liodakis}, {Maharana}, {Makrydopoulou}, {Mandarakas}, {Maragkakis},
  {Myserlis}, {Papadakis}, {Paterakis}, {Pearson}, {Ramaprakash}, {Readhead},
  {Reig}, {S{\l}owikowska}, {Tassis}, {Xexakis}, {{\.Z}ejmo}, \&
  {Zensus}}]{Blinov2021}
{Blinov}, D., {Kiehlmann}, S., {Pavlidou}, V., {et~al.} 2021, \mnras, 501, 3715

\bibitem[{{Blinov} {et~al.}(2023){Blinov}, {Maharana}, {Bouzelou}, {Casadio},
  {Gjerl{\o}w}, {Jormanainen}, {Kiehlmann}, {Kypriotakis}, {Liodakis},
  {Mandarakas}, {Markopoulioti}, {Panopoulou}, {Pelgrims}, {Pouliasi},
  {Romanopoulos}, {Skalidis}, {Anche}, {Angelakis}, {Antoniadis}, {Medhi},
  {Hovatta}, {Kus}, {Kylafis}, {Mahabal}, {Myserlis}, {Paleologou},
  {Papadakis}, {Pavlidou}, {Papamastorakis}, {Pearson}, {Potter},
  {Ramaprakash}, {Readhead}, {Reig}, {S{\l}owikowska}, {Tassis}, \&
  {Zensus}}]{Blinov2023}
{Blinov}, D., {Maharana}, S., {Bouzelou}, F., {et~al.} 2023, \aap, 677, A144

\bibitem[{{Blinov} {et~al.}(2018){Blinov}, {Pavlidou}, {Papadakis},
  {Kiehlmann}, {Liodakis}, {Panopoulou}, {Angelakis}, {Balokovi{\'c}},
  {Hovatta}, {King}, {Kus}, {Kylafis}, {Mahabal}, {Maharana}, {Myserlis},
  {Paleologou}, {Papamastorakis}, {Pazderski}, {Pearson}, {Ramaprakash},
  {Readhead}, {Reig}, {Tassis}, \& {Zensus}}]{Blinov2018}
{Blinov}, D., {Pavlidou}, V., {Papadakis}, I., {et~al.} 2018, \mnras, 474, 1296

\bibitem[{{Blinov} {et~al.}(2016{\natexlab{a}}){Blinov}, {Pavlidou},
  {Papadakis}, {Kiehlmann}, {Liodakis}, {Panopoulou}, {Pearson}, {Angelakis},
  {Balokovi{\'c}}, {Hovatta}, {Joshi}, {King}, {Kus}, {Kylafis}, {Mahabal},
  {Marecki}, {Myserlis}, {Paleologou}, {Papamastorakis}, {Pazderski},
  {Prabhudesai}, {Ramaprakash}, {Readhead}, {Reig}, {Tassis}, \&
  {Zensus}}]{Blinov2016-II}
{Blinov}, D., {Pavlidou}, V., {Papadakis}, I., {et~al.} 2016{\natexlab{a}},
  \mnras, 462, 1775

\bibitem[{{Blinov} {et~al.}(2015){Blinov}, {Pavlidou}, {Papadakis},
  {Kiehlmann}, {Panopoulou}, {Liodakis}, {King}, {Angelakis}, {Balokovi{\'c}},
  {Das}, {Feiler}, {Fuhrmann}, {Hovatta}, {Khodade}, {Kus}, {Kylafis},
  {Mahabal}, {Myserlis}, {Modi}, {Pazderska}, {Pazderski}, {Papamastorakis},
  {Pearson}, {Rajarshi}, {Ramaprakash}, {Reig}, {Readhead}, {Tassis}, \&
  {Zensus}}]{Blinov2015}
{Blinov}, D., {Pavlidou}, V., {Papadakis}, I., {et~al.} 2015, \mnras, 453, 1669

\bibitem[{{Blinov} {et~al.}(2016{\natexlab{b}}){Blinov}, {Pavlidou},
  {Papadakis}, {Hovatta}, {Pearson}, {Liodakis}, {Panopoulou}, {Angelakis},
  {Balokovi{\'c}}, {Das}, {Khodade}, {Kiehlmann}, {King}, {Kus}, {Kylafis},
  {Mahabal}, {Marecki}, {Modi}, {Myserlis}, {Paleologou}, {Papamastorakis},
  {Pazderska}, {Pazderski}, {Rajarshi}, {Ramaprakash}, {Readhead}, {Reig},
  {Tassis}, \& {Zensus}}]{Blinov2016}
{Blinov}, D., {Pavlidou}, V., {Papadakis}, I.~E., {et~al.} 2016{\natexlab{b}},
  \mnras, 457, 2252

\bibitem[{{Bowers} {et~al.}(2008){Bowers}, {Albright}, {Yin}, {Bergen}, \&
  {Kwan}}]{Bowers2008}
{Bowers}, K.~J., {Albright}, B.~J., {Yin}, L., {Bergen}, B., \& {Kwan},
  T.~J.~T. 2008, Physics of Plasmas, 15, 055703

\bibitem[{{Brown} {et~al.}(2013){Brown}, {Baliber}, {Bianco}, {Bowman},
  {Burleson}, {Conway}, {Crellin}, {Depagne}, {De Vera}, {Dilday}, {Dragomir},
  {Dubberley}, {Eastman}, {Elphick}, {Falarski}, {Foale}, {Ford}, {Fulton},
  {Garza}, {Gomez}, {Graham}, {Greene}, {Haldeman}, {Hawkins}, {Haworth},
  {Haynes}, {Hidas}, {Hjelstrom}, {Howell}, {Hygelund}, {Lister}, {Lobdill},
  {Martinez}, {Mullins}, {Norbury}, {Parrent}, {Paulson}, {Petry}, {Pickles},
  {Posner}, {Rosing}, {Ross}, {Sand}, {Saunders}, {Shobbrook}, {Shporer},
  {Street}, {Thomas}, {Tsapras}, {Tufts}, {Valenti}, {Vander Horst}, {Walker},
  {White}, \& {Willis}}]{LasCumbresObservatory2013}
{Brown}, T.~M., {Baliber}, N., {Bianco}, F.~B., {et~al.} 2013, \pasp, 125, 1031

\bibitem[{{Calafut} \& {Wiita}(2015)}]{Calafut2015}
{Calafut}, V. \& {Wiita}, P.~J. 2015, Journal of Astrophysics and Astronomy,
  36, 255

\bibitem[{{Chandra} {et~al.}(2014){Chandra}, {Baliyan}, {Ganesh}, \&
  {Foschini}}]{Chandra2014}
{Chandra}, S., {Baliyan}, K.~S., {Ganesh}, S., \& {Foschini}, L. 2014, \apj,
  791, 85

\bibitem[{{Chandra} {et~al.}(2012){Chandra}, {Baliyan}, {Ganesh}, \&
  {Joshi}}]{Chandra2012}
{Chandra}, S., {Baliyan}, K.~S., {Ganesh}, S., \& {Joshi}, U.~C. 2012, \apj,
  746, 92

\bibitem[{{de Jaeger} {et~al.}(2023){de Jaeger}, {Shappee}, {Kochanek},
  {Hinkle}, {Garrappa}, {Liodakis}, {Franckowiak}, {Stanek}, {Beacom}, \&
  {Prieto}}]{deJaeger2023}
{de Jaeger}, T., {Shappee}, B.~J., {Kochanek}, C.~S., {et~al.} 2023, \mnras,
  519, 6349

\bibitem[{{Di Gesu} {et~al.}(2022){Di Gesu}, {Donnarumma}, {Tavecchio},
  {Agudo}, {Barnounin}, {Cibrario}, {Di Lalla}, {Di Marco}, {Escudero},
  {Errando}, {Jorstad}, {Kim}, {Kouch}, {Liodakis}, {Lindfors}, {Madejski},
  {Marshall}, {Marscher}, {Middei}, {Muleri}, {Myserlis}, {Negro}, {Omodei},
  {Pacciani}, {Paggi}, {Perri}, {Puccetti}, {Antonelli}, {Bachetti}, {Baldini},
  {Baumgartner}, {Bellazzini}, {Bianchi}, {Bongiorno}, {Bonino}, {Brez},
  {Bucciantini}, {Capitanio}, {Castellano}, {Cavazzuti}, {Ciprini}, {Costa},
  {De Rosa}, {Del Monte}, {Doroshenko}, {Dov{\v{c}}iak}, {Ehlert}, {Enoto},
  {Evangelista}, {Fabiani}, {Ferrazzoli}, {Garcia}, {Gunji}, {Hayashida},
  {Heyl}, {Iwakiri}, {Karas}, {Kitaguchi}, {Kolodziejczak}, {Krawczynski}, {La
  Monaca}, {Latronico}, {Maldera}, {Manfreda}, {Marin}, {Marinucci}, {Massaro},
  {Matt}, {Mitsuishi}, {Mizuno}, {Ng}, {O'Dell}, {Oppedisano}, {Papitto},
  {Pavlov}, {Peirson}, {Pesce-Rollins}, {Petrucci}, {Pilia}, {Possenti},
  {Poutanen}, {Ramsey}, {Rankin}, {Ratheesh}, {Romani}, {Sgr{\`o}}, {Slane},
  {Soffitta}, {Spandre}, {Tamagawa}, {Taverna}, {Tawara}, {Tennant}, {Thomas},
  {Tombesi}, {Trois}, {Tsygankov}, {Turolla}, {Vink}, {Weisskopf}, {Wu}, {Xie},
  \& {Zane}}]{DiGesu2022}
{Di Gesu}, L., {Donnarumma}, I., {Tavecchio}, F., {et~al.} 2022, \apjl, 938, L7

\bibitem[{{Di Gesu} {et~al.}(2023){Di Gesu}, {Marshall}, {Ehlert}, {Kim},
  {Donnarumma}, {Tavecchio}, {Liodakis}, {Kiehlmann}, {Agudo}, {Jorstad},
  {Muleri}, {Marscher}, {Puccetti}, {Middei}, {Perri}, {Pacciani}, {Negro},
  {Romani}, {Di Marco}, {Blinov}, {Bourbah}, {Kontopodis}, {Mandarakas},
  {Romanopoulos}, {Skalidis}, {Vervelaki}, {Casadio}, {Escudero}, {Myserlis},
  {Gurwell}, {Rao}, {Keating}, {Kouch}, {Lindfors}, {Aceituno}, {Bernardos},
  {Bonnoli}, {Casanova}, {Garc{\'\i}a-Comas}, {Ag{\'\i}s-Gonz{\'a}lez},
  {Husillos}, {Marchini}, {Sota}, {Imazawa}, {Sasada}, {Fukazawa}, {Kawabata},
  {Uemura}, {Mizuno}, {Nakaoka}, {Akitaya}, {Savchenko}, {Vasilyev},
  {G{\'o}mez}, {Antonelli}, {Barnouin}, {Bonino}, {Cavazzuti}, {Costamante},
  {Chen}, {Cibrario}, {De Rosa}, {Di Pierro}, {Errando}, {Kaaret}, {Karas},
  {Krawczynski}, {Lisalda}, {Madejski}, {Malacaria}, {Marin}, {Marinucci},
  {Massaro}, {Matt}, {Mitsuishi}, {O'Dell}, {Paggi}, {Peirson}, {Petrucci},
  {Ramsey}, {Tennant}, {Wu}, {Bachetti}, {Baldini}, {Baumgartner},
  {Bellazzini}, {Bianchi}, {Bongiorno}, {Brez}, {Bucciantini}, {Capitanio},
  {Castellano}, {Ciprini}, {Costa}, {Del Monte}, {Di Lalla}, {Doroshenko},
  {Dov{\v{c}}iak}, {Enoto}, {Evangelista}, {Fabiani}, {Ferrazzoli}, {Garcia},
  {Gunji}, {Hayashida}, {Heyl}, {Iwakiri}, {Kislat}, {Kitaguchi},
  {Kolodziejczak}, {La Monaca}, {Latronico}, {Maldera}, {Manfreda}, {Ng},
  {Omodei}, {Oppedisano}, {Papitto}, {Pavlov}, {Pesce-Rollins}, {Pilia},
  {Possenti}, {Poutanen}, {Rankin}, {Ratheesh}, {Roberts}, {Sgr{\`o}}, {Slane},
  {Soffitta}, {Spandre}, {Swartz}, {Tamagawa}, {Taverna}, {Tawara}, {Thomas},
  {Tombesi}, {Trois}, {Tsygankov}, {Turolla}, {Vink}, {Weisskopf}, {Xie}, \&
  {Zane}}]{DiGesu2023}
{Di Gesu}, L., {Marshall}, H.~L., {Ehlert}, S.~R., {et~al.} 2023, Nature
  Astronomy, 7, 1245

\bibitem[{{Eisenstein} \& {Hut}(1998)}]{Eisenstein1998}
{Eisenstein}, D.~J. \& {Hut}, P. 1998, \apj, 498, 137

\bibitem[{{Giannios} {et~al.}(2009){Giannios}, {Uzdensky}, \&
  {Begelman}}]{Giannios2009}
{Giannios}, D., {Uzdensky}, D.~A., \& {Begelman}, M.~C. 2009, \mnras, 395, L29

\bibitem[{{Hodge} {et~al.}(2018){Hodge}, {Lister}, {Aller}, {Aller}, {Kovalev},
  {Pushkarev}, \& {Savolainen}}]{Hodge2018}
{Hodge}, M.~A., {Lister}, M.~L., {Aller}, M.~F., {et~al.} 2018, \apj, 862, 151

\bibitem[{{Hosking} \& {Sironi}(2020)}]{Hosking2020}
{Hosking}, D.~N. \& {Sironi}, L. 2020, \apjl, 900, L23

\bibitem[{{Hovatta} \& {Lindfors}(2019)}]{Hovatta2019}
{Hovatta}, T. \& {Lindfors}, E. 2019, \nar, 87, 101541

\bibitem[{{Hovatta} {et~al.}(2016){Hovatta}, {Lindfors}, {Blinov}, {Pavlidou},
  {Nilsson}, {Kiehlmann}, {Angelakis}, {Fallah Ramazani}, {Liodakis},
  {Myserlis}, {Panopoulou}, \& {Pursimo}}]{Hovatta2016}
{Hovatta}, T., {Lindfors}, E., {Blinov}, D., {et~al.} 2016, \aap, 596, A78

\bibitem[{{Hovatta} {et~al.}(2021){Hovatta}, {Lindfors}, {Kiehlmann},
  {Max-Moerbeck}, {Hodges}, {Liodakis}, {L{\"a}hteem{\"a}ki}, {Pearson},
  {Readhead}, {Reeves}, {Suutarinen}, {Tammi}, \& {Tornikoski}}]{Hovatta2021}
{Hovatta}, T., {Lindfors}, E., {Kiehlmann}, S., {et~al.} 2021, \aap, 650, A83

\bibitem[{{Ikejiri} {et~al.}(2011){Ikejiri}, {Uemura}, {Sasada}, {Ito},
  {Yamanaka}, {Sakimoto}, {Arai}, {Fukazawa}, {Ohsugi}, {Kawabata}, {Yoshida},
  {Sato}, \& {Kino}}]{Ikejiri2011}
{Ikejiri}, Y., {Uemura}, M., {Sasada}, M., {et~al.} 2011, \pasj, 63, 639

\bibitem[{{Imazawa} {et~al.}(2023){Imazawa}, {Sasada}, {Hazama}, {Fukazawa},
  {Kawabata}, {Nakaoka}, {Akitaya}, {Bohn}, \& {Gangopadhyay}}]{Imazawa2023}
{Imazawa}, R., {Sasada}, M., {Hazama}, N., {et~al.} 2023, \pasj, 75, 1

\bibitem[{{Jermak} {et~al.}(2016){Jermak}, {Steele}, {Lindfors}, {Hovatta},
  {Nilsson}, {Lamb}, {Mundell}, {Barres de Almeida}, {Berdyugin}, {Kadenius},
  {Reinthal}, \& {Takalo}}]{Jermak2016}
{Jermak}, H., {Steele}, I.~A., {Lindfors}, E., {et~al.} 2016, \mnras, 462, 4267

\bibitem[{{Jones} {et~al.}(1974){Jones}, {O'Dell}, \& {Stein}}]{Jones1974}
{Jones}, T.~W., {O'Dell}, S.~L., \& {Stein}, W.~A. 1974, \apj, 188, 353

\bibitem[{{Jormanainen} {et~al.}(2023){Jormanainen}, {Hovatta}, {Christie},
  {Lindfors}, {Petropoulou}, \& {Liodakis}}]{Jormanainen2023}
{Jormanainen}, J., {Hovatta}, T., {Christie}, I.~M., {et~al.} 2023, \aap, 678,
  A140

\bibitem[{{Jorstad} {et~al.}(2010){Jorstad}, {Marscher}, {Larionov}, {Agudo},
  {Smith}, {Gurwell}, {L{\"a}hteenm{\"a}ki}, {Tornikoski}, {Markowitz},
  {Arkharov}, {Blinov}, {Chatterjee}, {D'Arcangelo}, {Falcone}, {G{\'o}mez},
  {Hagen-Thorn}, {Jordan}, {Kimeridze}, {Konstantinova}, {Kopatskaya},
  {Kurtanidze}, {Larionova}, {Larionova}, {McHardy}, {Melnichuk},
  {Roca-Sogorb}, {Schmidt}, {Skiff}, {Taylor}, {Thum}, {Troitsky}, \&
  {Wiesemeyer}}]{Jorstad2010}
{Jorstad}, S.~G., {Marscher}, A.~P., {Larionov}, V.~M., {et~al.} 2010, \apj,
  715, 362

\bibitem[{{Jorstad} {et~al.}(2022){Jorstad}, {Marscher}, {Raiteri}, {Villata},
  {Weaver}, {Zhang}, {Dong}, {G{\'o}mez}, {Perel}, {Savchenko}, {Larionov},
  {Carosati}, {Chen}, {Kurtanidze}, {Marchini}, {Matsumoto}, {Mortari},
  {Aceti}, {Acosta-Pulido}, {Andreeva}, {Apolonio}, {Arena}, {Arkharov},
  {Bachev}, {Banfi}, {Bonnoli}, {Borman}, {Bozhilov}, {Carnerero},
  {Damljanovic}, {Ehgamberdiev}, {Els{\"a}sser}, {Frasca}, {Gabellini},
  {Grishina}, {Gupta}, {Hagen-Thorn}, {Hallum}, {Hart}, {Hasuda}, {Hemrich},
  {Hsiao}, {Ibryamov}, {Irsmambetova}, {Ivanov}, {Joner}, {Kimeridze},
  {Klimanov}, {Kn{\"o}tt}, {Kopatskaya}, {Kurtanidze}, {Kurtenkov}, {Kuutma},
  {Larionova}, {Leonini}, {Lin}, {Lorey}, {Mannheim}, {Marino}, {Minev},
  {Mirzaqulov}, {Morozova}, {Nikiforova}, {Nikolashvili}, {Ovcharov}, {Papini},
  {Pursimo}, {Rahimov}, {Reinhart}, {Sakamoto}, {Salvaggio}, {Semkov},
  {Shakhovskoy}, {Sigua}, {Steineke}, {Stojanovic}, {Strigachev}, {Troitskaya},
  {Troitskiy}, {Tsai}, {Valcheva}, {Vasilyev}, {Vince}, {Waller}, {Zaharieva},
  \& {Chatterjee}}]{Jorstad2022}
{Jorstad}, S.~G., {Marscher}, A.~P., {Raiteri}, C.~M., {et~al.} 2022, \nat,
  609, 265

\bibitem[{{Kiehlmann}(2015)}]{2015PhDT.......630K}
{Kiehlmann}, S. 2015, PhD thesis, Andreas Eckart University of Cologne, Germany

\bibitem[{{Kiehlmann}(2024)}]{Kiehlmann2024}
{Kiehlmann}, S. 2024, {polarizationtools: Polarization analysis and simulation
  tools in python}, Astrophysics Source Code Library, record ascl:2402.006

\bibitem[{{Kiehlmann} {et~al.}(2021){Kiehlmann}, {Blinov}, {Liodakis},
  {Pavlidou}, {Readhead}, {Angelakis}, {Casadio}, {Hovatta}, {Kylafis},
  {Mahabal}, {Mandarakas}, {Myserlis}, {Panopoulou}, {Pearson}, {Ramaprakash},
  {Reig}, {Skalidis}, {S{\l}owikowska}, {Tassis}, \& {Zensus}}]{Kiehlmann2021}
{Kiehlmann}, S., {Blinov}, D., {Liodakis}, I., {et~al.} 2021, \mnras, 507, 225

\bibitem[{{Kiehlmann} {et~al.}(2016){Kiehlmann}, {Savolainen}, {Jorstad},
  {Sokolovsky}, {Schinzel}, {Marscher}, {Larionov}, {Agudo}, {Akitaya},
  {Ben{\'\i}tez}, {Berdyugin}, {Blinov}, {Bochkarev}, {Borman}, {Burenkov},
  {Casadio}, {Doroshenko}, {Efimova}, {Fukazawa}, {G{\'o}mez}, {Grishina},
  {Hagen-Thorn}, {Heidt}, {Hiriart}, {Itoh}, {Joshi}, {Kawabata}, {Kimeridze},
  {Kopatskaya}, {Korobtsev}, {Krajci}, {Kurtanidze}, {Kurtanidze}, {Larionova},
  {Larionova}, {Lindfors}, {L{\'o}pez}, {McHardy}, {Molina}, {Moritani},
  {Morozova}, {Nazarov}, {Nikolashvili}, {Nilsson}, {Pulatova}, {Reinthal},
  {Sadun}, {Sasada}, {Savchenko}, {Sergeev}, {Sigua}, {Smith}, {Sorcia},
  {Spiridonova}, {Takaki}, {Takalo}, {Taylor}, {Troitsky}, {Uemura},
  {Ugolkova}, {Ui}, {Yoshida}, {Zensus}, \& {Zhdanova}}]{Kiehlmann2016}
{Kiehlmann}, S., {Savolainen}, T., {Jorstad}, S.~G., {et~al.} 2016, \aap, 590,
  A10

\bibitem[{{Kim} {et~al.}(2024){Kim}, {Di Gesu}, {Liodakis}, {Marscher},
  {Jorstad}, {Middei}, {Marshall}, {Pacciani}, {Agudo}, {Tavecchio},
  {Cibrario}, {Tugliani}, {Bonino}, {Negro}, {Puccetti}, {Tombesi}, {Costa},
  {Donnarumma}, {Soffitta}, {Mizuno}, {Fukazawa}, {Kawabata}, {Nakaoka},
  {Uemura}, {Imazawa}, {Sasada}, {Akitaya}, {Jos{\`e} Aceituno}, {Bonnoli},
  {Casanova}, {Myserlis}, {Sievers}, {Angelakis}, {Kraus}, {Yeon Cheong},
  {Jeong}, {Kang}, {Kim}, {Lee}, {Ag{\`\i}s-Gonz{\`a}lez}, {Sota}, {Escudero},
  {Gurwell}, {Keating}, {Rao}, {Kouch}, {Lindfors}, {Bourbah}, {Kiehlmann},
  {Kontopodis}, {Mandarakas}, {Romanopoulos}, {Skalidis}, {Vervelaki},
  {Savchenko}, {Antonelli}, {Bachetti}, {Baldini}, {Baumgartner}, {Bellazzini},
  {Bianchi}, {Bongiorno}, {Brez}, {Bucciantini}, {Capitanio}, {Castellano},
  {Cavazzuti}, {Chen}, {Ciprini}, {De Rosa}, {Del Monte}, {Di Lalla}, {Di
  Marco}, {Doroshenko}, {Dov{\v{c}}iak}, {Ehlert}, {Enoto}, {Evangelista},
  {Fabiani}, {Ferrazzoli}, {Garcia}, {Gunji}, {Hayashida}, {Heyl}, {Iwakiri},
  {Kaaret}, {Karas}, {Kislat}, {Kitaguchi}, {Kolodziejczak}, {Krawczynski}, {La
  Monaca}, {Latronico}, {Maldera}, {Manfreda}, {Marin}, {Marinucci}, {Massaro},
  {Matt}, {Mitsuishi}, {Muleri}, {Ng}, {O'Dell}, {Omodei}, {Oppedisano},
  {Papitto}, {Pavlov}, {Peirson}, {Perri}, {Pesce-Rollins}, {Petrucci},
  {Pilia}, {Possenti}, {Poutanen}, {Ramsey}, {Rankin}, {Ratheesh}, {Roberts},
  {Romani}, {Sgr{\'o}}, {Slane}, {Spandre}, {Swartz}, {Tamagawa}, {Taverna},
  {Tawara}, {Tennant}, {Thomas}, {Trois}, {Tsygankov}, {Turolla}, {Vink},
  {Weisskopf}, {Wu}, {Xie}, \& {Zane}}]{Kim2024}
{Kim}, D.~E., {Di Gesu}, L., {Liodakis}, I., {et~al.} 2024, \aap, 681, A12

\bibitem[{{King} {et~al.}(2014){King}, {Blinov}, {Ramaprakash}, {Myserlis},
  {Angelakis}, {Balokovi{\'c}}, {Feiler}, {Fuhrmann}, {Hovatta}, {Khodade},
  {Kougentakis}, {Kylafis}, {Kus}, {Modi}, {Paleologou}, {Panopoulou},
  {Papadakis}, {Papamastorakis}, {Paterakis}, {Pavlidou}, {Pazderska},
  {Pazderski}, {Pearson}, {Rajarshi}, {Readhead}, {Reig}, {Steiakaki},
  {Tassis}, \& {Zensus}}]{King2014}
{King}, O.~G., {Blinov}, D., {Ramaprakash}, A.~N., {et~al.} 2014, \mnras, 442,
  1706

\bibitem[{{Komarov} {et~al.}(2020){Komarov}, {Moskvitin}, {Bychkov},
  {Burenkov}, {Drabek}, {Shergin}, {Emelyanov}, {Komarova}, {Romanenko}, \&
  {Aitov}}]{2020AstBu..75..486K}
{Komarov}, V.~V., {Moskvitin}, A.~S., {Bychkov}, V.~D., {et~al.} 2020,
  Astrophysical Bulletin, 75, 486

\bibitem[{{Kouch} {et~al.}(2024){Kouch}, {Liodakis}, {Middei}, {Kim},
  {Tavecchio}, {Marscher}, {Marshall}, {Ehlert}, {Di Gesu}, {Jorstad}, {Agudo},
  {Madejski}, {Romani}, {Errando}, {Lindfors}, {Nilsson}, {Toppari}, {Potter},
  {Imazawa}, {Sasada}, {Fukazawa}, {Kawabata}, {Uemura}, {Mizuno}, {Nakaoka},
  {Akitaya}, {McCall}, {Jermak}, {Steele}, {Myserlis}, {Gurwell}, {Keating},
  {Rao}, {Kang}, {Lee}, {Kim}, {Cheong}, {Jeong}, {Angelakis}, {Kraus},
  {Jos{\'e} Aceituno}, {Bonnoli}, {Casanova}, {Escudero},
  {Ag{\'\i}s-Gonz{\'a}lez}, {Husillos}, {Morcuende}, {Otero-Santos}, {Sota},
  {Bachev}, {Antonelli}, {Bachetti}, {Baldini}, {Baumgartner}, {Bellazzini},
  {Bianchi}, {Bongiorno}, {Bonino}, {Brez}, {Bucciantini}, {Capitanio},
  {Castellano}, {Cavazzuti}, {Chen}, {Ciprini}, {Costa}, {De Rosa}, {Del
  Monte}, {Di Lalla}, {Di Marco}, {Donnarumma}, {Doroshenko}, {Dov{\v{c}}iak},
  {Enoto}, {Evangelista}, {Fabiani}, {Ferrazzoli}, {Garcia}, {Gunji},
  {Hayashida}, {Heyl}, {Iwakiri}, {Kaaret}, {Karas}, {Kislat}, {Kitaguchi},
  {Kolodziejczak}, {Krawczynski}, {La Monaca}, {Latronico}, {Maldera},
  {Manfreda}, {Marin}, {Marinucci}, {Massaro}, {Matt}, {Mitsuishi}, {Muleri},
  {Negro}, {Ng}, {O'Dell}, {Omodei}, {Oppedisano}, {Papitto}, {Pavlov},
  {Peirson}, {Perri}, {Pesce-Rollins}, {Petrucci}, {Pilia}, {Possenti},
  {Poutanen}, {Puccetti}, {Ramsey}, {Rankin}, {Ratheesh}, {Roberts},
  {Sgr{\`o}}, {Slane}, {Soffitta}, {Spandre}, {Swartz}, {Tamagawa}, {Taverna},
  {Tawara}, {Tennant}, {Thomas}, {Tombesi}, {Trois}, {Tsygankov}, {Turolla},
  {Vink}, {Weisskopf}, {Wu}, {Xie}, \& {Zane}}]{Kouch2024}
{Kouch}, P.~M., {Liodakis}, I., {Middei}, R., {et~al.} 2024, arXiv e-prints,
  arXiv:2406.01693

\bibitem[{{Larionov} {et~al.}(2008){Larionov}, {Jorstad}, {Marscher},
  {Raiteri}, {Villata}, {Agudo}, {Aller}, {Arkharov}, {Asfandiyarov}, {Bach},
  {Bachev}, {Berdyugin}, {B{\"o}ttcher}, {Buemi}, {Calcidese}, {Carosati},
  {Charlot}, {Chen}, {di Paola}, {Dolci}, {Dogru}, {Doroshenko}, {Efimov},
  {Erdem}, {Frasca}, {Fuhrmann}, {Giommi}, {Glowienka}, {Gupta}, {Gurwell},
  {Hagen-Thorn}, {Hsiao}, {Ibrahimov}, {Jordan}, {Kamada}, {Konstantinova},
  {Kopatskaya}, {Kovalev}, {Kovalev}, {Kurtanidze}, {L{\"a}hteenm{\"a}ki},
  {Lanteri}, {Larionova}, {Leto}, {Le Campion}, {Lee}, {Lindfors}, {Marilli},
  {McHardy}, {Mingaliev}, {Nazarov}, {Nieppola}, {Nilsson}, {Ohlert},
  {Pasanen}, {Porter}, {Pursimo}, {Ros}, {Sadakane}, {Sadun}, {Sergeev},
  {Smith}, {Strigachev}, {Sumitomo}, {Takalo}, {Tanaka}, {Trigilio}, {Umana},
  {Ungerechts}, {Volvach}, \& {Yuan}}]{Larionov2008}
{Larionov}, V.~M., {Jorstad}, S.~G., {Marscher}, A.~P., {et~al.} 2008, \aap,
  492, 389

\bibitem[{Liodakis(2024)}]{DVN/IETSXS_2024}
Liodakis, I. 2024, {NOPE observations of BL Lacertae and CGRaBS J0211+1051}

\bibitem[{{Liodakis} {et~al.}(2020){Liodakis}, {Blinov}, {Jorstad}, {Arkharov},
  {Di Paola}, {Efimova}, {Grishina}, {Kiehlmann}, {Kopatskaya}, {Larionov},
  {Larionova}, {Larionova}, {Marscher}, {Morozova}, {Nikiforova}, {Pavlidou},
  {Traianou}, {Troitskaya}, {Troitsky}, {Uemura}, \& {Weaver}}]{Liodakis2020}
{Liodakis}, I., {Blinov}, D., {Jorstad}, S.~G., {et~al.} 2020, \apj, 902, 61

\bibitem[{{Liodakis} {et~al.}(2022){Liodakis}, {Marscher}, {Agudo},
  {Berdyugin}, {Bernardos}, {Bonnoli}, {Borman}, {Casadio}, {Casanova},
  {Cavazzuti}, {Rodriguez Cavero}, {Di Gesu}, {Di Lalla}, {Donnarumma},
  {Ehlert}, {Errando}, {Escudero}, {Garc{\'\i}a-Comas},
  {Ag{\'\i}s-Gonz{\'a}lez}, {Husillos}, {Jormanainen}, {Jorstad}, {Kagitani},
  {Kopatskaya}, {Kravtsov}, {Krawczynski}, {Lindfors}, {Larionova}, {Madejski},
  {Marin}, {Marchini}, {Marshall}, {Morozova}, {Massaro}, {Masiero}, {Mawet},
  {Middei}, {Millar-Blanchaer}, {Myserlis}, {Negro}, {Nilsson}, {O'Dell},
  {Omodei}, {Pacciani}, {Paggi}, {Panopoulou}, {Peirson}, {Perri}, {Petrucci},
  {Poutanen}, {Puccetti}, {Romani}, {Sakanoi}, {Savchenko}, {Sota},
  {Tavecchio}, {Tinyanont}, {Vasilyev}, {Weaver}, {Zhovtan}, {Antonelli},
  {Bachetti}, {Baldini}, {Baumgartner}, {Bellazzini}, {Bianchi}, {Bongiorno},
  {Bonino}, {Brez}, {Bucciantini}, {Capitanio}, {Castellano}, {Ciprini},
  {Costa}, {De Rosa}, {Del Monte}, {Di Marco}, {Doroshenko}, {Dov{\v{c}}iak},
  {Enoto}, {Evangelista}, {Fabiani}, {Ferrazzoli}, {Garcia}, {Gunji},
  {Hayashida}, {Heyl}, {Iwakiri}, {Karas}, {Kitaguchi}, {Kolodziejczak}, {La
  Monaca}, {Latronico}, {Maldera}, {Manfreda}, {Marinucci}, {Matt},
  {Mitsuishi}, {Mizuno}, {Muleri}, {Ng}, {Oppedisano}, {Papitto}, {Pavlov},
  {Pesce-Rollins}, {Pilia}, {Possenti}, {Ramsey}, {Rankin}, {Ratheesh},
  {Sgr{\'o}}, {Slane}, {Soffitta}, {Spandre}, {Tamagawa}, {Taverna}, {Tawara},
  {Tennant}, {Thomas}, {Tombesi}, {Trois}, {Tsygankov}, {Turolla}, {Vink},
  {Weisskopf}, {Wu}, {Xie}, \& {Zane}}]{Liodakis2022}
{Liodakis}, I., {Marscher}, A.~P., {Agudo}, I., {et~al.} 2022, \nat, 611, 677

\bibitem[{{Liodakis} {et~al.}(2019{\natexlab{a}}){Liodakis}, {Peirson}, \&
  {Romani}}]{Liodakis2019-II}
{Liodakis}, I., {Peirson}, A.~L., \& {Romani}, R.~W. 2019{\natexlab{a}}, \apj,
  880, 29

\bibitem[{{Liodakis} {et~al.}(2018){Liodakis}, {Romani}, {Filippenko},
  {Kiehlmann}, {Max-Moerbeck}, {Readhead}, \& {Zheng}}]{Liodakis2018}
{Liodakis}, I., {Romani}, R.~W., {Filippenko}, A.~V., {et~al.} 2018, \mnras,
  480, 5517

\bibitem[{{Liodakis} {et~al.}(2019{\natexlab{b}}){Liodakis}, {Romani},
  {Filippenko}, {Kocevski}, \& {Zheng}}]{Liodakis2019}
{Liodakis}, I., {Romani}, R.~W., {Filippenko}, A.~V., {Kocevski}, D., \&
  {Zheng}, W. 2019{\natexlab{b}}, \apj, 880, 32

\bibitem[{{Lister} {et~al.}(2021){Lister}, {Homan}, {Kellermann}, {Kovalev},
  {Pushkarev}, {Ros}, \& {Savolainen}}]{Lister2021}
{Lister}, M.~L., {Homan}, D.~C., {Kellermann}, K.~I., {et~al.} 2021, \apj, 923,
  30

\bibitem[{{MAGIC Collaboration} {et~al.}(2018){MAGIC Collaboration}, {Ahnen},
  {Ansoldi}, {Antonelli}, {Arcaro}, {Baack}, {Babi{\'c}}, {Banerjee},
  {Bangale}, {Barres de Almeida}, {Barrio}, {Becerra Gonz{\'a}lez}, {Bednarek},
  {Bernardini}, {Ch Berse}, {Berti}, {Bhattacharyya}, {Biland}, {Blanch},
  {Bonnoli}, {Carosi}, {Carosi}, {Ceribella}, {Chatterjee}, {Colak}, {Colin},
  {Colombo}, {Contreras}, {Cortina}, {Covino}, {Cumani}, {da Vela}, {Dazzi},
  {de Angelis}, {de Lotto}, {Delfino}, {Delgado}, {di Pierro},
  {Dom{\'\i}nguez}, {Dominis Prester}, {Dorner}, {Doro}, {Einecke},
  {Elsaesser}, {Fallah Ramazani}, {Fern{\'a}ndez-Barral}, {Fidalgo}, {Fonseca},
  {Font}, {Fruck}, {Galindo}, {Gallozzi}, {Garc{\'\i}a L{\'o}pez},
  {Garczarczyk}, {Gaug}, {Giammaria}, {Godinovi{\'c}}, {Gora}, {Guberman},
  {Hadasch}, {Hahn}, {Hassan}, {Hayashida}, {Herrera}, {Hose}, {Hrupec},
  {Ishio}, {Konno}, {Kubo}, {Kushida}, {Kuve{\v{z}}di{\'c}}, {Lelas},
  {Lindfors}, {Lombardi}, {Longo}, {L{\'o}pez}, {Maggio}, {Majumdar},
  {Makariev}, {Maneva}, {Manganaro}, {Mannheim}, {Maraschi}, {Mariotti},
  {Mart{\'\i}nez}, {Masuda}, {Mazin}, {Mielke}, {Minev}, {Miranda}, {Mirzoyan},
  {Moralejo}, {Moreno}, {Moretti}, {Nagayoshi}, {Neustroev}, {Niedzwiecki},
  {Nievas Rosillo}, {Nigro}, {Nilsson}, {Ninci}, {Nishijima}, {Noda},
  {Nogu{\'e}s}, {Paiano}, {Palacio}, {Paneque}, {Paoletti}, {Paredes},
  {Pedaletti}, {Peresano}, {Persic}, {Prada Moroni}, {Prandini}, {Puljak},
  {Garcia}, {Reichardt}, {Rhode}, {Rib{\'o}}, {Rico}, {Righi}, {Rugliancich},
  {Saito}, {Satalecka}, {Schweizer}, {Sitarek}, {{\v{S}}nidari{\'c}},
  {Sobczynska}, {Stamerra}, {Strzys}, {Suri{\'c}}, {Takahashi}, {Takalo},
  {Tavecchio}, {Temnikov}, {Terzi{\'c}}, {Teshima}, {Torres-Alb{\`a}},
  {Treves}, {Tsujimoto}, {Vanzo}, {Vazquez Acosta}, {Vovk}, {Ward}, {Will},
  {Zari{\'c}}, {Fermi-Lat Collaboration}, {Bastieri}, {Gasparrini}, {Lott},
  {Rani}, {Thompson}, {MWL Collaborators}, {Agudo}, {Angelakis}, {Borman},
  {Casadio}, {Grishina}, {Gurwell}, {Hovatta}, {Itoh}, {J{\"a}rvel{\"a}},
  {Jermak}, {Jorstad}, {Kopatskaya}, {Kraus}, {Krichbaum}, {Kuin},
  {L{\"a}hteenm{\"a}ki}, {Larionov}, {Larionova}, {Lien}, {Madejski},
  {Marscher}, {Myserlis}, {Max-Moerbeck}, {Molina}, {Morozova}, {Nalewajko},
  {Pearson}, {Ramakrishnan}, {Readhead}, {Reeves}, {Savchenko}, {Steele},
  {Tornikoski}, {Troitskaya}, {Troitsky}, {Vasilyev}, \&
  {Zensus}}]{MAGICCollaboration2018}
{MAGIC Collaboration}, {Ahnen}, M.~L., {Ansoldi}, S., {et~al.} 2018, \aap, 619,
  A45

\bibitem[{{Marscher}(2014)}]{Marscher2014}
{Marscher}, A.~P. 2014, \apj, 780, 87

\bibitem[{{Marscher} \& {Gear}(1985)}]{Marscher1985}
{Marscher}, A.~P. \& {Gear}, W.~K. 1985, \apj, 298, 114

\bibitem[{{Marscher} \& {Jorstad}(2021)}]{Marscher2021}
{Marscher}, A.~P. \& {Jorstad}, S.~G. 2021, Galaxies, 9, 27

\bibitem[{{Marscher} \& {Jorstad}(2022)}]{Marscher2022}
{Marscher}, A.~P. \& {Jorstad}, S.~G. 2022, Universe, 8, 644

\bibitem[{{Marscher} {et~al.}(2008){Marscher}, {Jorstad}, {D'Arcangelo},
  {Smith}, {Williams}, {Larionov}, {Oh}, {Olmstead}, {Aller}, {Aller},
  {McHardy}, {L{\"a}hteenm{\"a}ki}, {Tornikoski}, {Valtaoja}, {Hagen-Thorn},
  {Kopatskaya}, {Gear}, {Tosti}, {Kurtanidze}, {Nikolashvili}, {Sigua},
  {Miller}, \& {Ryle}}]{Marscher2008}
{Marscher}, A.~P., {Jorstad}, S.~G., {D'Arcangelo}, F.~D., {et~al.} 2008, \nat,
  452, 966

\bibitem[{{Marscher} {et~al.}(2010){Marscher}, {Jorstad}, {Larionov}, {Aller},
  {Aller}, {L{\"a}hteenm{\"a}ki}, {Agudo}, {Smith}, {Gurwell}, {Hagen-Thorn},
  {Konstantinova}, {Larionova}, {Larionova}, {Melnichuk}, {Blinov},
  {Kopatskaya}, {Troitsky}, {Tornikoski}, {Hovatta}, {Schmidt}, {D'Arcangelo},
  {Bhattarai}, {Taylor}, {Olmstead}, {Manne-Nicholas}, {Roca-Sogorb},
  {G{\'o}mez}, {McHardy}, {Kurtanidze}, {Nikolashvili}, {Kimeridze}, \&
  {Sigua}}]{Marscher2010}
{Marscher}, A.~P., {Jorstad}, S.~G., {Larionov}, V.~M., {et~al.} 2010, \apjl,
  710, L126

\bibitem[{{Mead} {et~al.}(1990){Mead}, {Ballard}, {Brand}, {Hough}, {Brindle},
  \& {Bailey}}]{Mead1990}
{Mead}, A.~R.~G., {Ballard}, K.~R., {Brand}, P.~W.~J.~L., {et~al.} 1990, \aaps,
  83, 183

\bibitem[{{Meisner} \& {Romani}(2010)}]{Meisner2010}
{Meisner}, A.~M. \& {Romani}, R.~W. 2010, \apj, 712, 14

\bibitem[{{Meyer} {et~al.}(2019){Meyer}, {Scargle}, \& {Blandford}}]{Meyer2019}
{Meyer}, M., {Scargle}, J.~D., \& {Blandford}, R.~D. 2019, \apj, 877, 39

\bibitem[{{Middei} {et~al.}(2023){Middei}, {Liodakis}, {Perri}, {Puccetti},
  {Cavazzuti}, {Di Gesu}, {Ehlert}, {Madejski}, {Marscher}, {Marshall},
  {Muleri}, {Negro}, {Jorstad}, {Ag{\'\i}s-Gonz{\'a}lez}, {Agudo}, {Bonnoli},
  {Bernardos}, {Casanova}, {Garc{\'\i}a-Comas}, {Husillos}, {Marchini}, {Sota},
  {Kouch}, {Lindfors}, {Borman}, {Kopatskaya}, {Larionova}, {Morozova},
  {Savchenko}, {Vasilyev}, {Zhovtan}, {Casadio}, {Escudero}, {Myserlis},
  {Hales}, {Kameno}, {Kneissl}, {Messias}, {Nagai}, {Blinov}, {Bourbah},
  {Kiehlmann}, {Kontopodis}, {Mandarakas}, {Romanopoulos}, {Skalidis},
  {Vervelaki}, {Masiero}, {Mawet}, {Millar-Blanchaer}, {Panopoulou},
  {Tinyanont}, {Berdyugin}, {Kagitani}, {Kravtsov}, {Sakanoi}, {Imazawa},
  {Sasada}, {Fukazawa}, {Kawabata}, {Uemura}, {Mizuno}, {Nakaoka}, {Akitaya},
  {Gurwell}, {Rao}, {Di Lalla}, {Cibrario}, {Donnarumma}, {Kim}, {Omodei},
  {Pacciani}, {Poutanen}, {Tavecchio}, {Antonelli}, {Bachetti}, {Baldini},
  {Baumgartner}, {Bellazzini}, {Bianchi}, {Bongiorno}, {Bonino}, {Brez},
  {Bucciantini}, {Capitanio}, {Castellano}, {Ciprini}, {Costa}, {De Rosa}, {Del
  Monte}, {Di Marco}, {Doroshenko}, {Dov{\v{c}}iak}, {Enoto}, {Evangelista},
  {Fabiani}, {Ferrazzoli}, {Garcia}, {Gunji}, {Hayashida}, {Heyl}, {Iwakiri},
  {Karas}, {Kitaguchi}, {Kolodziejczak}, {Krawczynski}, {La Monaca},
  {Latronico}, {Maldera}, {Manfreda}, {Marin}, {Marinucci}, {Massaro}, {Matt},
  {Mitsuishi}, {Ng}, {O'Dell}, {Oppedisano}, {Papitto}, {Pavlov}, {Peirson},
  {Pesce-Rollins}, {Petrucci}, {Pilia}, {Possenti}, {Ramsey}, {Rankin},
  {Ratheesh}, {Romani}, {Sgr{\'o}}, {Slane}, {Soffitta}, {Spandre}, {Tamagawa},
  {Taverna}, {Tawara}, {Tennant}, {Thomas}, {Tombesi}, {Trois}, {Tsygankov},
  {Turolla}, {Vink}, {Weisskopf}, {Wu}, {Xie}, \& {Zane}}]{Middei2023}
{Middei}, R., {Liodakis}, I., {Perri}, M., {et~al.} 2023, \apjl, 942, L10

\bibitem[{{Morozova} {et~al.}(2014){Morozova}, {Larionov}, {Troitsky},
  {Jorstad}, {Marscher}, {G{\'o}mez}, {Blinov}, {Efimova}, {Hagen-Thorn},
  {Hagen-Thorn}, {Joshi}, {Konstantinova}, {Kopatskaya}, {Larionova},
  {Larionova}, {L{\"a}hteenm{\"a}ki}, {Tammi}, {Rastorgueva-Foi}, {McHardy},
  {Tornikoski}, {Agudo}, {Casadio}, {Molina}, {Volvach}, \&
  {Volvach}}]{Morozova2014}
{Morozova}, D.~A., {Larionov}, V.~M., {Troitsky}, I.~S., {et~al.} 2014, \aj,
  148, 42

\bibitem[{{Nilsson} {et~al.}(2018){Nilsson}, {Lindfors}, {Takalo}, {Reinthal},
  {Berdyugin}, {Sillanp{\"a}{\"a}}, {Ciprini}, {Halkola}, {Hein{\"a}m{\"a}ki},
  {Hovatta}, {Kadenius}, {Nurmi}, {Ostorero}, {Pasanen}, {Rekola}, {Saarinen},
  {Sainio}, {Tuominen}, {Villforth}, {Vornanen}, \& {Zaprudin}}]{Nilsson2018}
{Nilsson}, K., {Lindfors}, E., {Takalo}, L.~O., {et~al.} 2018, \aap, 620, A185

\bibitem[{{Panopoulou} {et~al.}(2015){Panopoulou}, {Tassis}, {Blinov},
  {Pavlidou}, {King}, {Paleologou}, {Ramaprakash}, {Angelakis},
  {Balokovi{\'c}}, {Das}, {Feiler}, {Hovatta}, {Khodade}, {Kiehlmann}, {Kus},
  {Kylafis}, {Liodakis}, {Mahabal}, {Modi}, {Myserlis}, {Papadakis},
  {Papamastorakis}, {Pazderska}, {Pazderski}, {Pearson}, {Rajarshi},
  {Readhead}, {Reig}, \& {Zensus}}]{Panopoulou2015}
{Panopoulou}, G., {Tassis}, K., {Blinov}, D., {et~al.} 2015, \mnras, 452, 715

\bibitem[{{Peirson} {et~al.}(2022){Peirson}, {Liodakis}, \&
  {Romani}}]{Peirson2022}
{Peirson}, A.~L., {Liodakis}, I., \& {Romani}, R.~W. 2022, \apj, 931, 59

\bibitem[{{Peirson} {et~al.}(2023){Peirson}, {Negro}, {Liodakis}, {Middei},
  {Kim}, {Marscher}, {Marshall}, {Pacciani}, {Romani}, {Wu}, {Di Marco}, {Di
  Lalla}, {Omodei}, {Jorstad}, {Agudo}, {Kouch}, {Lindfors}, {Aceituno},
  {Bernardos}, {Bonnoli}, {Casanova}, {Garc{\'\i}a-Comas},
  {Ag{\'\i}s-Gonz{\'a}lez}, {Husillos}, {Marchini}, {Sota}, {Casadio},
  {Escudero}, {Myserlis}, {Sievers}, {Gurwell}, {Rao}, {Imazawa}, {Sasada},
  {Fukazawa}, {Kawabata}, {Uemura}, {Mizuno}, {Nakaoka}, {Akitaya}, {Cheong},
  {Jeong}, {Kang}, {Kim}, {Lee}, {Angelakis}, {Kraus}, {Cibrario},
  {Donnarumma}, {Poutanen}, {Tavecchio}, {Antonelli}, {Bachetti}, {Baldini},
  {Baumgartner}, {Bellazzini}, {Bianchi}, {Bongiorno}, {Bonino}, {Brez},
  {Bucciantini}, {Capitanio}, {Castellano}, {Cavazzuti}, {Chen}, {Ciprini},
  {Costa}, {De Rosa}, {Del Monte}, {Di Gesu}, {Doroshenko}, {Dov{\v{c}}iak},
  {Ehlert}, {Enoto}, {Evangelista}, {Fabiani}, {Ferrazzoli}, {Garcia}, {Gunji},
  {Hayashida}, {Heyl}, {Iwakiri}, {Kaaret}, {Karas}, {Kitaguchi},
  {Kolodziejczak}, {Krawczynski}, {La Monaca}, {Latronico}, {Madejski},
  {Maldera}, {Manfreda}, {Marin}, {Marinucci}, {Massaro}, {Matt}, {Mitsuishi},
  {Muleri}, {Ng}, {O'Dell}, {Oppedisano}, {Papitto}, {Pavlov}, {Perri},
  {Pesce-Rollins}, {Petrucci}, {Pilia}, {Possenti}, {Puccetti}, {Ramsey},
  {Rankin}, {Ratheesh}, {Roberts}, {Sgr{\'o}}, {Slane}, {Soffitta}, {Spandre},
  {Swartz}, {Tamagawa}, {Taverna}, {Tawara}, {Tennant}, {Thomas}, {Tombesi},
  {Trois}, {Tsygankov}, {Turolla}, {Vink}, {Weisskopf}, {Xie}, \&
  {Zane}}]{Peirson2023}
{Peirson}, A.~L., {Negro}, M., {Liodakis}, I., {et~al.} 2023, \apjl, 948, L25

\bibitem[{{Peirson} \& {Romani}(2018)}]{Peirson2018}
{Peirson}, A.~L. \& {Romani}, R.~W. 2018, \apj, 864, 140

\bibitem[{{Peirson} \& {Romani}(2019)}]{Peirson2019}
{Peirson}, A.~L. \& {Romani}, R.~W. 2019, \apj, 885, 76

\bibitem[{{Piirola} {et~al.}(2014){Piirola}, {Berdyugin}, \&
  {Berdyugina}}]{Piirola2014}
{Piirola}, V., {Berdyugin}, A., \& {Berdyugina}, S. 2014, in \procspie, Vol.
  9147, Ground-based and Airborne Instrumentation for Astronomy V, 91478I

\bibitem[{{Piirola} {et~al.}(2020){Piirola}, {Berdyugin}, {Frisch}, {Kagitani},
  {Sakanoi}, {Berdyugina}, {Cole}, {Harlingten}, \& {Hill}}]{Piirola2020}
{Piirola}, V., {Berdyugin}, A., {Frisch}, P.~C., {et~al.} 2020, \aap, 635, A46

\bibitem[{{Raiteri} {et~al.}(2017{\natexlab{a}}){Raiteri}, {Nicastro},
  {Stamerra}, {Villata}, {Larionov}, {Blinov}, {Acosta-Pulido}, {Ar{\'e}valo},
  {Arkharov}, {Bachev}, {Borman}, {Carnerero}, {Carosati}, {Cecconi}, {Chen},
  {Damljanovic}, {Di Paola}, {Ehgamberdiev}, {Frasca}, {Giroletti},
  {Gonz{\'a}lez-Morales}, {Gri{\~n}on-Mar{\'\i}n}, {Grishina}, {Huang},
  {Ibryamov}, {Klimanov}, {Kopatskaya}, {Kurtanidze}, {Kurtanidze},
  {L{\"a}hteenm{\"a}ki}, {Larionova}, {Larionova}, {L{\'a}zaro}, {Leto},
  {Liodakis}, {Mart{\'\i}nez-Lombilla}, {Mihov}, {Mirzaqulov}, {Mokrushina},
  {Moody}, {Morozova}, {Nazarov}, {Nikolashvili}, {Ohlert}, {Panopoulou},
  {Pastor Yabar}, {Pinna}, {Protasio}, {Rizzi}, {Sadun}, {Savchenko}, {Semkov},
  {Sigua}, {Slavcheva-Mihova}, {Strigachev}, {Tornikoski}, {Troitskaya},
  {Troitsky}, {Vasilyev}, {Vera}, {Vince}, \& {Zanmar
  Sanchez}}]{Raiteri2017-II}
{Raiteri}, C.~M., {Nicastro}, F., {Stamerra}, A., {et~al.} 2017{\natexlab{a}},
  \mnras, 466, 3762

\bibitem[{{Raiteri} \& {Villata}(2021)}]{Raiteri2021-II}
{Raiteri}, C.~M. \& {Villata}, M. 2021, Galaxies, 9, 42

\bibitem[{{Raiteri} {et~al.}(2017{\natexlab{b}}){Raiteri}, {Villata},
  {Acosta-Pulido}, {Agudo}, {Arkharov}, {Bachev}, {Baida}, {Ben{\'\i}tez},
  {Borman}, {Boschin}, {Bozhilov}, {Butuzova}, {Calcidese}, {Carnerero},
  {Carosati}, {Casadio}, {Castro-Segura}, {Chen}, {Damljanovic}, {D'Ammando},
  {di Paola}, {Echevarr{\'\i}a}, {Efimova}, {Ehgamberdiev}, {Espinosa},
  {Fuentes}, {Giunta}, {G{\'o}mez}, {Grishina}, {Gurwell}, {Hiriart}, {Jermak},
  {Jordan}, {Jorstad}, {Joshi}, {Kopatskaya}, {Kuratov}, {Kurtanidze},
  {Kurtanidze}, {L{\"a}hteenm{\"a}ki}, {Larionov}, {Larionova}, {Larionova},
  {L{\'a}zaro}, {Lin}, {Malmrose}, {Marscher}, {Matsumoto}, {McBreen},
  {Michel}, {Mihov}, {Minev}, {Mirzaqulov}, {Mokrushina}, {Molina}, {Moody},
  {Morozova}, {Nazarov}, {Nikolashvili}, {Ohlert}, {Okhmat}, {Ovcharov},
  {Pinna}, {Polakis}, {Protasio}, {Pursimo}, {Redondo-Lorenzo}, {Rizzi},
  {Rodriguez-Coira}, {Sadakane}, {Sadun}, {Samal}, {Savchenko}, {Semkov},
  {Skiff}, {Slavcheva-Mihova}, {Smith}, {Steele}, {Strigachev}, {Tammi},
  {Thum}, {Tornikoski}, {Troitskaya}, {Troitsky}, {Vasilyev}, \&
  {Vince}}]{Raiteri2017}
{Raiteri}, C.~M., {Villata}, M., {Acosta-Pulido}, J.~A., {et~al.}
  2017{\natexlab{b}}, \nat, 552, 374

\bibitem[{{Raiteri} {et~al.}(2023){Raiteri}, {Villata}, {Jorstad}, {Marscher},
  {Acosta Pulido}, {Carosati}, {Chen}, {Joner}, {Kurtanidze}, {Lorey},
  {Marchini}, {Matsumoto}, {Mirzaqulov}, {Savchenko}, {Strigachev}, {Vince},
  {Aceti}, {Apolonio}, {Arena}, {Arkharov}, {Bachev}, {Bader}, {Banfi},
  {Bonnoli}, {Borman}, {Bozhilov}, {Brown}, {Carbonell}, {Carnerero},
  {Damljanovic}, {Dhiman}, {Ehgamberdiev}, {Elsaesser}, {Feige}, {Gabellini},
  {Gal{\'a}n}, {Galli}, {Gaur}, {Gazeas}, {Grishina}, {Gupta}, {Hagen-Thorn},
  {Hallum}, {Hart}, {Hasuda}, {Heidemann}, {Horst}, {Hou}, {Ibryamov},
  {Ivanidze}, {Jovanovic}, {Kimeridze}, {Kishore}, {Klimanov}, {Kopatskaya},
  {Kurtanidze}, {Kushwaha}, {Lane}, {Larionova}, {Leonini}, {Lin}, {Mannheim},
  {Marino}, {Minev}, {Modaressi}, {Morozova}, {Mortari}, {Nazarov},
  {Nikolashvili}, {Otero Santos}, {Ovcharov}, {Papini}, {Pinter}, {Privitera},
  {Pursimo}, {Reinhart}, {Roberts}, {Romanov}, {Rosenlehner}, {Sakamoto},
  {Salvaggio}, {Schoch}, {Semkov}, {Seufert}, {Shakhovskoy}, {Sigua}, {Singh},
  {Steineke}, {Stojanovic}, {Tripathi}, {Troitskaya}, {Troitskiy}, {Tsai},
  {Valcheva}, {Vasilyev}, {Vrontaki}, {Weaver}, {Wooley}, {Zaharieva}, \&
  {Zhovtan}}]{Raiteri2023}
{Raiteri}, C.~M., {Villata}, M., {Jorstad}, S.~G., {et~al.} 2023, \mnras, 522,
  102

\bibitem[{{Raiteri} {et~al.}(2021){Raiteri}, {Villata}, {Larionov}, {Jorstad},
  {Marscher}, {Weaver}, {Acosta-Pulido}, {Agudo}, {Andreeva}, {Arkharov},
  {Bachev}, {Ben{\'\i}tez}, {Berton}, {Bj{\"o}rklund}, {Borman}, {Bozhilov},
  {Carnerero}, {Carosati}, {Casadio}, {Chen}, {Damljanovic}, {D'Ammando},
  {Escudero}, {Fuentes}, {Giroletti}, {Grishina}, {Gupta}, {Hagen-Thorn},
  {Hart}, {Hiriart}, {Hou}, {Ivanov}, {Kim}, {Kimeridze}, {Konstantopoulou},
  {Kopatskaya}, {Kurtanidze}, {Kurtanidze}, {L{\"a}hteenm{\"a}ki}, {Larionova},
  {Larionova}, {Marchili}, {Markovic}, {Minev}, {Morozova}, {Myserlis},
  {Nakamura}, {Nikiforova}, {Nikolashvili}, {Otero-Santos}, {Ovcharov},
  {Pursimo}, {Rahimov}, {Righini}, {Sakamoto}, {Savchenko}, {Semkov},
  {Shakhovskoy}, {Sigua}, {Stojanovic}, {Strigachev}, {Thum}, {Tornikoski},
  {Traianou}, {Troitskaya}, {Troitskiy}, {Tsai}, {Valcheva}, {Vasilyev},
  {Vince}, \& {Zaharieva}}]{Raiteri2021}
{Raiteri}, C.~M., {Villata}, M., {Larionov}, V.~M., {et~al.} 2021, \mnras, 504,
  5629

\bibitem[{{Ramaprakash} {et~al.}(2019){Ramaprakash}, {Rajarshi}, {Das},
  {Khodade}, {Modi}, {Panopoulou}, {Maharana}, {Blinov}, {Angelakis},
  {Casadio}, {Fuhrmann}, {Hovatta}, {Kiehlmann}, {King}, {Kylafis},
  {Kougentakis}, {Kus}, {Mahabal}, {Marecki}, {Myserlis}, {Paterakis},
  {Paleologou}, {Liodakis}, {Papadakis}, {Papamastorakis}, {Pavlidou},
  {Pazderski}, {Pearson}, {Readhead}, {Reig}, {S{\l}owikowska}, {Tassis}, \&
  {Zensus}}]{Ramaprakash2019}
{Ramaprakash}, A.~N., {Rajarshi}, C.~V., {Das}, H.~K., {et~al.} 2019, \mnras,
  485, 2355

\bibitem[{{Sasada} {et~al.}(2017){Sasada}, {Mineshige}, {Yamada}, \&
  {Negoro}}]{Sasada2017}
{Sasada}, M., {Mineshige}, S., {Yamada}, S., \& {Negoro}, H. 2017, \pasj, 69,
  15

\bibitem[{{Scargle} {et~al.}(2013){Scargle}, {Norris}, {Jackson}, \&
  {Chiang}}]{Scargle2013}
{Scargle}, J.~D., {Norris}, J.~P., {Jackson}, B., \& {Chiang}, J. 2013, \apj,
  764, 167

\bibitem[{{Shao} {et~al.}(2019){Shao}, {Jiang}, \& {Chen}}]{Shao2019}
{Shao}, X., {Jiang}, Y., \& {Chen}, X. 2019, \apj, 884, 15

\bibitem[{{Steele} {et~al.}(2004){Steele}, {Smith}, {Rees}, {Baker}, {Bates},
  {Bode}, {Bowman}, {Carter}, {Etherton}, {Ford}, {Fraser}, {Gomboc}, {Lett},
  {Mansfield}, {Marchant}, {Medrano-Cerda}, {Mottram}, {Raback}, {Scott},
  {Tomlinson}, \& {Zamanov}}]{Steele2004}
{Steele}, I.~A., {Smith}, R.~J., {Rees}, P.~C., {et~al.} 2004, in Proceedings
  of SPIE, Vol. 5489, Ground-based Telescopes, ed. J.~{Oschmann}, Jacobus~M.,
  679--692

\bibitem[{{Tavecchio}(2021)}]{Tavecchio2021}
{Tavecchio}, F. 2021, Galaxies, 9, 37

\bibitem[{{Uemura} {et~al.}(2017){Uemura}, {Itoh}, {Liodakis}, {Blinov},
  {Nakayama}, {Xu}, {Sawada}, {Wu}, \& {Fujishiro}}]{Uemura2017}
{Uemura}, M., {Itoh}, R., {Liodakis}, I., {et~al.} 2017, \pasj, 69, 96

\bibitem[{{Weaver} {et~al.}(2022){Weaver}, {Jorstad}, {Marscher}, {Morozova},
  {Troitsky}, {Agudo}, {G{\'o}mez}, {L{\"a}hteenm{\"a}ki}, {Tammi}, \&
  {Tornikoski}}]{Weaver2022}
{Weaver}, Z.~R., {Jorstad}, S.~G., {Marscher}, A.~P., {et~al.} 2022, \apjs,
  260, 12

\bibitem[{{Weaver} {et~al.}(2020){Weaver}, {Williamson}, {Jorstad}, {Marscher},
  {Larionov}, {Raiteri}, {Villata}, {Acosta-Pulido}, {Bachev}, {Baida},
  {Balonek}, {Ben{\'\i}tez}, {Borman}, {Bozhilov}, {Carnerero}, {Carosati},
  {Chen}, {Damljanovic}, {Dhiman}, {Dougherty}, {Ehgamberdiev}, {Grishina},
  {Gupta}, {Hart}, {Hiriart}, {Hsiao}, {Ibryamov}, {Joner}, {Kimeridze},
  {Kopatskaya}, {Kurtanidze}, {Kurtanidze}, {Larionova}, {Matsumoto},
  {Matsumura}, {Minev}, {Mirzaqulov}, {Morozova}, {Nikiforova}, {Nikolashvili},
  {Ovcharov}, {Rizzi}, {Sadun}, {Savchenko}, {Semkov}, {Slater}, {Smith},
  {Stojanovic}, {Strigachev}, {Troitskaya}, {Troitsky}, {Tsai}, {Vince},
  {Valcheva}, {Vasilyev}, {Zaharieva}, \& {Zhovtan}}]{Weaver2020}
{Weaver}, Z.~R., {Williamson}, K.~E., {Jorstad}, S.~G., {et~al.} 2020, \apj,
  900, 137

\bibitem[{{Webb} \& {Sanz}(2023)}]{Webb2023}
{Webb}, J.~R. \& {Sanz}, I.~P. 2023, Galaxies, 11, 108

\bibitem[{{Weisskopf} {et~al.}(2022){Weisskopf}, {Soffitta}, {Baldini},
  {Ramsey}, {O'Dell}, {Romani}, {Matt}, {Deininger}, {Baumgartner},
  {Bellazzini}, {Costa}, {Kolodziejczak}, {Latronico}, {Marshall}, {Muleri},
  {Bongiorno}, {Tennant}, {Bucciantini}, {Dovciak}, {Marin}, {Marscher},
  {Poutanen}, {Slane}, {Turolla}, {Kalinowski}, {Di Marco}, {Fabiani},
  {Minuti}, {La Monaca}, {Pinchera}, {Rankin}, {Sgro'}, {Trois}, {Xie},
  {Alexander}, {Allen}, {Amici}, {Andersen}, {Antonelli}, {Antoniak},
  {Attin{\`a}}, {Barbanera}, {Bachetti}, {Baggett}, {Bladt}, {Brez}, {Bonino},
  {Boree}, {Borotto}, {Breeding}, {Brienza}, {Bygott}, {Caporale}, {Cardelli},
  {Carpentiero}, {Castellano}, {Castronuovo}, {Cavalli}, {Cavazzuti},
  {Ceccanti}, {Centrone}, {Citraro}, {D'Amico}, {D'Alba}, {Di Gesu}, {Del
  Monte}, {Dietz}, {Di Lalla}, {Persio}, {Dolan}, {Donnarumma}, {Evangelista},
  {Ferrant}, {Ferrazzoli}, {Ferrie}, {Footdale}, {Forsyth}, {Foster},
  {Garelick}, {Gunji}, {Gurnee}, {Head}, {Hibbard}, {Johnson}, {Kelly},
  {Kilaru}, {Lefevre}, {Roy}, {Loffredo}, {Lorenzi}, {Lucchesi}, {Maddox},
  {Magazzu}, {Maldera}, {Manfreda}, {Mangraviti}, {Marengo}, {Marrocchesi},
  {Massaro}, {Mauger}, {McCracken}, {McEachen}, {Mize}, {Mereu}, {Mitchell},
  {Mitsuishi}, {Morbidini}, {Mosti}, {Nasimi}, {Negri}, {Negro}, {Nguyen},
  {Nitschke}, {Nuti}, {Onizuka}, {Oppedisano}, {Orsini}, {Osborne}, {Pacheco},
  {Paggi}, {Painter}, {Pavelitz}, {Pentz}, {Piazzolla}, {Perri},
  {Pesce-Rollins}, {Peterson}, {Pilia}, {Profeti}, {Puccetti}, {Ranganathan},
  {Ratheesh}, {Reedy}, {Root}, {Rubini}, {Ruswick}, {Sanchez}, {Sarra},
  {Santoli}, {Scalise}, {Sciortino}, {Schroeder}, {Seek}, {Sosdian}, {Spandre},
  {Speegle}, {Tamagawa}, {Tardiola}, {Tobia}, {Thomas}, {Valerie}, {Vimercati},
  {Walden}, {Weddendorf}, {Wedmore}, {Welch}, {Zanetti}, \&
  {Zanetti}}]{Weisskopf2022}
{Weisskopf}, M.~C., {Soffitta}, P., {Baldini}, L., {et~al.} 2022, JATIS, 8,
  026002

\bibitem[{{Wiersema} {et~al.}(2023){Wiersema}, {Starling}, {Campagnolo},
  {Thanki}, \& {McErlean}}]{Wiersema2023}
{Wiersema}, K., {Starling}, R.~L.~C., {Campagnolo}, J.~C.~N., {Thanki}, D., \&
  {McErlean}, R. 2023, RAS Techniques and Instruments, 2, 106

\bibitem[{{Zhang} {et~al.}(2014){Zhang}, {Chen}, \& {B{\"o}ttcher}}]{Zhang2014}
{Zhang}, H., {Chen}, X., \& {B{\"o}ttcher}, M. 2014, \apj, 789, 66

\bibitem[{{Zhang} {et~al.}(2017){Zhang}, {Li}, {Guo}, \& {Taylor}}]{Zhang2017}
{Zhang}, H., {Li}, H., {Guo}, F., \& {Taylor}, G. 2017, \apj, 835, 125

\bibitem[{{Zhang} {et~al.}(2020){Zhang}, {Li}, {Giannios}, {Guo}, {Liu}, \&
  {Dong}}]{Zhang2020}
{Zhang}, H., {Li}, X., {Giannios}, D., {et~al.} 2020, \apj, 901, 149

\end{thebibliography}

\end{document}